\documentstyle[12pt,epsf]{article}
\input{psfig}

\makeatletter

\@addtoreset{equation}{section}
\makeatother
\newcommand{\be}{\begin{equation}}
\newcommand{\ee}{\end{equation}}
\newcommand{\ba}{\begin{eqnarray}}
\newcommand{\ea}{\end{eqnarray}}
\newcommand{\dle}[1]{\label{#1}}
\newcommand{\dla}[1]{\label{#1}}
\newcommand{\dr}[1]{\ref{#1}}
\newcommand{\dc}[1]{\cite{#1}}

\newcommand{\gsim}{\raise.3ex\hbox{$>$\kern-.75em\lower1ex\hbox{$\sim$}}}
\newcommand{\lsim}{\raise.3ex\hbox{$<$\kern-.75em\lower1ex\hbox{$\sim$}}}

\newcommand{\k}{\ell}
\newcommand{\dk}{d\k}
\newcommand{\kp}{{\ell}'}

\newcommand{\nlt}{n(\ell,t)}

\newcommand{\gzt}{g(z,t)}
\newcommand{\gzpt}{g(z',t)}
\newcommand{\paa}{\partial}
\newcommand{\dl}{\dot{\ell}}
\newcommand{\Ai}{A(z;z',z-z')}
\newcommand{\Aiii}{A(z+z';z,z')}
\newcommand{\Bii}{B(z,z'-z;z')}
\newcommand{\Biiii}{B(z',z-z';z)}
\newcommand{\bxi}{\bar{\xi}}
\newcommand{\bxic}{\bar{\xi}_c}
\newcommand{\zetac}{\zeta_c}
\newcommand{\tchi}{\tilde{\chi}}
\newcommand{\GGmu}{\Gamma G \mu}
\newcommand{\half}{{\frac{1}{2}}}

\begin{document}

\typeout{--- Title page start ---}

\renewcommand{\thefootnote}{\fnsymbol{footnote}}

\begin{flushright} SUSX-TH-98-005 \\
Imperial/TP/97-98/31 \\
DAMTP-97-152 \\
\end{flushright}
\vskip 25pt

\begin{center} {\large\bf The evolution of a network of cosmic
string loops}\\
\vskip 1.2cm {\large  E.J.Copeland$^a$\footnote{E-mail: {\tt
e.j.copeland@sussex.ac.uk}},
T.W.B.Kibble$^b$\footnote{E-mail: {\tt t.kibble@ic.ac.uk}} and
D.A.Steer$^{c}$\footnote{E-mail: {\tt D.A.Steer@damtp.cam.ac.uk}}}\\
\vskip 5pt
{\it a}) Centre for Theoretical Physics, University of
Sussex, \\
Brighton BN1 9QH, U.K.\\
\vskip 3pt
 {\it b}) Blackett
Laboratory, Imperial College, Prince Consort Road,\\ London SW7 2BZ,
U.K.\\
\vskip 3pt
{\it c}) D.A.M.T.P., Silver Street, Cambridge, CB3 9EW, U.K.\\
\vskip 6pt
\today
\end{center}

\vskip 1cm

\begin{abstract}
We set up and analyse a model for the non-equilibrium evolution of a
network of cosmic strings initially containing {\em only loops} and
{\em no infinite strings}.  Due to this particular initial condition, our
analytical approach differs significantly from existing ones.  We
describe the average properties of the network in terms of the
distribution function $\nlt \dk$, the average number of loops per unit
volume with physical length between
$\k$ and $\k + \dk$ at time $t$.  The dynamical processes which change
the length of loops are then estimated and an equation, which we call
the `rate equation', is derived for
$\paa \nlt/\paa t$.  In a non-expanding universe, the loops should
reach the equilibrium distribution predicted by string statistical
mechanics.  Analysis of the rate equation gives results consistent
with this.  We then study the rate equation in an expanding universe
and suggest that
three different final states are possible for the
evolving loop network, each of which may well be realised for some
initial conditions.  If the initial energy
density in loops in the radiation era is low, then the loops
rapidly disappear.
For large initial energy densities, we expect that either infinite strings
are formed or that the loops tend towards a
scaling solution in the radiation era and then rapidly disappear in
the matter era.  Such a scenario may be relevant given recent work
highlighting the problems with structure formation from the standard
cosmic string scenario \dc{James}.

\end{abstract}

\vskip 1cm

\renewcommand{\thefootnote}{\arabic{footnote}}
\setcounter{footnote}{0}
\typeout{--- Main Text Start ---}

\newpage
\section{Introduction}

Cosmic strings are just one of a number of different topological
defects which may have formed at a phase transition very early in
the history of the universe \dc{Tom,HiKi,ViSh}.   Since they are
topologically stable, cosmic strings may have
survived until today and thus may provide
an alternative theory to inflation for explaining the
formation of structures such as galaxies in the universe, and the
temperature fluctuations in the cosmic microwave background (CMBR).
Over the last twenty years, a `standard' picture
has emerged of the initial distribution of strings produced at the
phase transition:  numerical simulations of
string formation
\dc{VV} suggest that this distribution
consists of two distinct `populations'.  The first are
cosmic string {\em loops} (strings which are closed on themselves) whose
length distribution is smooth, typically a scale
invariant $\k^{-5/2}$
power-law.  The second is a separate population of {\em infinite}
strings (we define the difference between a long loop and an
infinite string below).  Numerical simulations
\dc{AT,BB,AS} and
analytic models \dc{ACK,CJAP} of the subsequent
evolution of such a network in a critical expanding universe
then suggest that the strings
reach a `scaling' solution in which the energy density in
strings is a fixed fraction of the energy density of the universe and
all scales in the problem grow as the horizon.
Given this `standard' picture, calculations of the observational
effects of cosmic strings
together with comparison with experimental data suggest that
GUT scale strings {\em may not} be responsible for both the 
fluctuations in the
CMBR as well as the density perturbations \dc{James}.
These authors suggest that strings or other active perturbations might
be reconciled with
data if they could somehow disappear rapidly at a redshift of about
100.  Alternatively one could consider
the effects of strings in a universe with a cosmological
constant \dc{James2,Avelino}.

However, one should recall that many very important
details are still unknown both about the formation and
evolution of cosmic strings.  For example, there is uncertainty
in the order of early universe phase transitions \dc{Kaj}:  can
topological defects
form at all in continuous transitions?  If they can, what would their
initial distribution be and would it still contain the two different
populations?  Even in first and second order transitions, analytic
verification of the numerical results of string formation is very
difficult given the non-equilibrium and non-perturbative nature of the
problem \dc{RayGlk}.  Recently attention has once again turned to
understanding the effects of the lattice
\dc{KarlHi,Julian,ScVi,AndyJames,AndyTom,MarkGray,Vach}.
There are also obvious limitations to numerical simulations
of the evolution of cosmic string networks. For instance, because of the
complexity of the system, these simulations can only run for a small 
fraction
($\sim 10^{-25}$) of the total time between string formation and today.
Also important gravitational back reaction effects have yet to be
incorporated in numerical work and any string structure which is
smaller than the
lattice size cannot be resolved.  These limitations mean, for example,
that the small scale structure on strings is not treated as accurately
as it needs to be.
Analytical models for the evolution of a network of cosmic strings
containing infinite strings have also been developed.  One such model
has been proposed by two of us \dc{ACK}, and in it we included the
effects of back reaction.  The production
of loops from infinite strings was also treated, but we
ignored the subsequent fate of the
loops assuming that they all decayed by gravitational radiation.

In this paper we develop a rather different analytical model for
the evolution of a network of cosmic strings that contains {\em only}
loops.  Before outlining some of our motivation for considering this
problem, we define what we mean by infinite strings and loops.  The
reason for doing this is that there might be some confusion about the
difference between very long loops (whose length tends to infinity)
and infinite strings.  Also, most
definitions of infinite strings are tailor made to strings on lattices
with periodic boundary conditions \dc{VV}.  Here we work instead in an
infinite
space --- an $\Omega = 1$ universe --- and hence there are no boundary
conditions.   To define the meaning of an infinite string, consider a
cube of side $X$, and let $n(\k)$ be the
number density of closed strings with length $\k$ which we assume to
be Brownian random walks of step $\bxi$.  (We want to show
that by definition $n(\k)$ must only describe a loop distribution
without infinite strings.)  Since the length density ${L} =
\int_{0}^{\infty} \k n(\k) d\k$ in strings
must be finite, then if $n(\k) \propto \k^{- \alpha}$ for large $\k$
this implies that $\alpha >
2$.  Within the cube, the total length in strings
which {\em do not} touch the walls is
$$ {\bf L}_{in} = \int_0^{X^2/\bxi} \left(X - \sqrt{\k \bxi} 
\right)^3 \k n(\k)
\dk. $$
By definition, the total length of closed strings in the volume is
given by
$$ {\bf L}_{tot}= X^3 \int_0^{\infty} n(\k) \k \dk. $$
If, in the limit $X \rightarrow \infty$, the ratio $
{\bf L}_{in}/{\bf L}_{tot} \rightarrow 1$ then we define this to mean that
there are only loops and no infinite strings; that is, as $X
\rightarrow \infty$ no strings are long
enough to cross the
boundary.  Since this ratio is indeed 1 for all $n(\k)$ for which ${L}$
is finite,
it follows that the distribution $n(\k)$ does
describe solely loops.  Infinite strings take the form of a
separate distribution which is characterised by an $X$ independent length
density $L_{inf}$; this just adds a contribution to
$ {\bf L}_{tot}$ so that typically the ratio ${\bf L}_{in}/{\bf L}_{tot}
\rightarrow
0.25$ \dc{VV,Ant2}.  For any $X$, working with only a loop
distribution means that there are less strings which cross the box
than there would be if a population of infinite strings were also present.

Our motivation for considering the evolution of such a network of
loops comes from
different directions.  Firstly, it is interesting to focus
on the loop network alone (this, we believe, has never been done), and
hence try to understand its properties better.  For example, it is
often said that infinite strings are responsible for seeding the
perturbations which then collapse to give galaxies.  However, could
a loop network not survive long
enough to perform the same task?  Existing numerical work suggests that
the evolution of
a loop network may differ
significantly from that of one that also contains infinite strings
\dc{Pedro}.
Loop networks may therefore
have very different cosmological consequences.  In particular,
might the evolution be such as to `save' cosmic strings --- i.e.\
that the loops scale in the radiation era and then disappear in the
matter era \dc{James}?  Finally, as a
result of a very different dynamical
simulation of a first order phase transition, Borrill \dc{Julian}
has argued that
there is no evidence for infinite strings, but only a population of
loops.  As yet
this claim has not been disproved even though it has been
contested \dc{ScVi,AndyJames,Vach}.  It is certainly the case
that the fraction of infinite strings decreases if there is a variation in
domain size at the phase transition \dc{AndyTom}.
Given our previous comments
on the complicated nature of phase transitions,
we do not believe
that an initial distribution containing only loops is ruled out for
certain.

In this paper, we therefore start from the hypothesis that
the initial distribution of cosmic strings formed at the phase
transition contains {\em only} loops and no infinite strings.
We set up a model
for the evolution of such a network and analyse
some of its properties.  Our model can be applied both to a loop
network evolving in a non-expanding universe, as well as to a network
in a critical $\Omega = 1$ expanding universe.  In the
non-expanding case, our dynamical approach gives results consistent
with those
of string statistical mechanics.  In an expanding universe
there are no results to compare with, and our goal is to determine
whether the loops ever reach a stable scaling solution.

This paper is organised in the following way.  Section \dr{model}
introduces the model and contains a
description of our approach.  We also specify how loops interact with
each other and derive an equation for the rate of
change with time of
$\nlt d\k$.  This equation is subsequently referred to as
the `rate equation'.  In section \dr{approx} we estimate some of the
coefficients which appear in it and specify our approximations and
assumptions.  In an expanding universe, it is
useful to change variables to `scaling' variables and this is done in
section \dr{sscalvar}.  Sections \dr{powerexp} and \dr{inf} are
concerned with the large $\k$ solutions of the rate equation, both
in a non-expanding as well as an expanding universe.  Questions of
stability and scaling are also discussed.  We summarise our
conclusions in section \dr{conc}.

\section{The Model}\dle{model}

In this section we set up a model to describe the non-equilibrium
evolution of a network of cosmic string loops.  Such
evolution may result from the expansion of the universe, or
alternatively it
could describe the evolution of the network towards an equilibrium
state in a non-expanding universe.
Throughout we use a statistical description of the loop network.

\subsection{Boltzmann equation for a loop network}\dle{bel}

The network of relativistic cosmic string loops is described by
the distribution function
\be
\nlt \dk ,
\dle{loopdist}
\ee
so that $\nlt \dk$ is the average number of loops per unit volume
with physical
length between
$\k$ and $\k+d\k$ at time $t$.  (For an analogy with a system of
particles, see
\dc{LL,Bern,KT}).  The dimensions of $n$ are thus given
by\footnote{We set $c=1$.}
\be
\left[ \nlt \dk \right] = \frac{1}{({\rm
Time})^3}.
\dle{loopdim}
\ee
Note that the
loop distribution $\nlt$ is taken to be homogeneous and isotropic in
space so that it does not depend on the position ${\bf r}$ of the
loop.\footnote{The
distribution function could depend on other parameters such as the
mean square velocity of the loops.  Here we consider the simplest case
and ignore all dependence other than that on $\k$.}

An equation for ${\paa \nlt }/{\paa t}$ can be obtained once
length-changing interactions between the
loops have been specified.  In order to do this, we assume throughout 
this work
that when two different loops
intersect, or when a loop self-intersects, partners are
exchanged \dc{HiKi,ViSh}.  Ignoring the effects of gravitational
radiation and expansion for the moment (see below), the two dominant
length-changing interactions between loops
are then
\begin{itemize}
\item A loop of length $\k$ can intersect another loop of length
$\k'$ to give a loop of length $\k+\k'$.
\item A loop of length $\k$ can self-intersect to produce two daughter
loops of length $\k'$ and $\k-\k'$.
\end{itemize}

Consider the first of these processes.  If the initial lengths of the
two loops
are in the range $\k \rightarrow \k+d\k$ and $\k'
\rightarrow \k'+d\k'$, then the number of collisions per unit time per unit
volume of this type is
\be
A(\k + \k';\k,\k')(n(\k,t)d\k) (n(\k',t) d\k').
\dle{AAdef}
\ee
As expected, this is proportional to the number of loops with each of the
initial lengths times the probability $A(\k + \k';\k,\k') $ of the
collision occurring.  Equation (\dr{AAdef}) defines $A$ which will
then 
be estimated in section \dr{approx}.
Dimensional analysis
gives
\be
\left[A(\k+\k';\k,\k') \right] = ({\rm Time})^2.
\dle{loopdim3}
\ee
Observe that by definition $A$ must be symmetric;
\be
A(\k+\k';\k,\k') = A(\k+\k';\k',\k),
\dle{symmA}
\ee
and that this process can both remove loops of length $\k$ and
produce them (as a result of the intersection of two smaller loops).

Similarly, the number of daughter loops of length $\k' $ produced
from the self-intersection of a loop of length $\k \rightarrow
\k+d\k$ per unit time per unit volume is given by
\be
{B}(\k-\k',\k';\k)(n(\k,t) d\k).
\dle{BBdef}
\ee
The function
${B}(\k-\k',\k';\k)$ defined by (\dr{BBdef}) has dimensions
\be
\left[{B}(\k-\k',\k';\k) \right] = \frac{1}{({\rm Time})^2}.
\dle{loopdim4}
\ee
Again note that $B$ is symmetric with
\be
{B}(\k-\k',\k';\k) = {B}(\k',\k-\k';\k).
\dle{Bsymm}
\ee
We will estimate both $A$ and $B$ in section \dr{approx}.

In a complete microscopical description of the system, these $A$ and
$B$ processes would be related though time reversal.  Here, however, we
are considering the average properties of the network and label
the loops only by their length.  Since this does not constitute a complete
set of state labels for the loops, we no longer expect $A$ and $B$
to be related in any simple way.

\subsection{Derivation of the rate equation}\dle{deriv}

An equation for ${\partial n}/{\partial
t}$ can be obtained in terms of
the functions
$A$ and $B$.  Recall first that in an expanding
universe, $\k$ is not a constant
between collisions, $\dl \neq 0$, since the length of the loops changes
for example though gravitational radiation \dc{HiKi,ViSh}.
Let $H$ be the expansion rate, and consider a
length interval from $\k$ to $\k + d\k$.  In a time $\Delta t$, the
change $\Delta(\nlt d\k)$ in the number of loops per
unit volume with length $\k
\rightarrow
\k + d\k$, is given by the number of loops per unit volume
which scatter into the interval minus the loops per unit volume
which scatter out.  Taking care with the special interval
from $0$ to $\dk$, the resulting
{\em rate equation} is obtained by
letting $\Delta t \rightarrow 0$ so that ${\Delta(\nlt
d\k)}/{\Delta t}
\rightarrow
({\paa n}/{\paa t})d\k$.  It is
\ba
\frac{\paa n}{\paa t}& = &
-\nlt  \frac{\paa {\dl}}{\paa \k} -
\frac{\paa \nlt }{\paa \k} {\dl} + \lim_{\k \rightarrow
0^+}\left[n(\k,t) \dot{\k}
\right] \delta(\k) - 3H\nlt
\dle{fulla}
 \\
 & +&
 \int_{0}^{\k/2}A(\k;\k',\k-\k')n(\k',t)n(\k-\k',t)d\kp \; \; \; \; \leftarrow
\mbox{Term 1}
\dle{full1}
 \\
& + &  \int_{\k}^{\infty} B(\k,\k'-\k;\k')n(\k',t)d\kp\; \; \; \; \; \;
\; \; \; \; \; \; \; \; \; \; \; \; \; \; \; \leftarrow
\mbox{Term 2}
\dle{full2}
 \\
& - & \nlt   \int_{0}^{\infty} A(\k+\k';\k,\k')n(\k',t)d\kp\; \; \; \;
\; \; \; \; \; \; \;  \leftarrow
\mbox{Term 3}
\dle{full3}
 \\
& - & \nlt   \int_{0}^{\k/2}  B(\k',\k-\k';\k)d\kp \; \; \; \; \; \;
\; \; \; \; \; \; \; \; \; \; \; \; \; \; \leftarrow
\mbox{Term 4}
\dle{fulleqn}.
\ea
In line
(\dr{fulla}), the first two terms can be written as the single
differential $-{\paa (n \dl)}/{\paa \k}$.  Hence they represent the
net positive flux of loops in $\k$ space.  The
third term in (\dr{fulla}) guarantees that
\be
\nlt = 0 \; \; \; \forall  \k < 0, \; \; \; \forall t,
\ee
as is required for a physical distribution of loops.  (For a
discussion of this point see also section \dr{sscondbeta}.)
The final term in (\dr{fulla})
arises from the expansion of the universe.

The four
scattering integrals (\dr{full1})-(\dr{fulleqn}) are subsequently
referred to as `term 1' to
`term 4' respectively. (Terms 3 and 4 are defined {\em without} the
minus signs.)
Term 1 gives the rate of production of loops of length $\k$ from the
intersection of two smaller loops of length $\k'$ and $\k-\k'$.
Notice that because of the symmetry of the $A$ coefficient
(\dr{symmA}), the upper limit in term 1 is $\k/2$ so as to avoid
over-counting.
Term 1 gives
a non-linear and non-local contribution to the rate
equation.   Term 3 is the
converse of term 1 and is the rate at which loops of length $\k$
disappear due to intersections with other loops.  Loops of length $\k$
may be produced as a result of a larger loop self-intersecting --- this
is term 2.  Its counterpart is term 4 which gives the
rate at which loops of length $\k$ disappear due to self-intersection.
Again the upper limit of term 4 is $\k/2$ so as to avoid
over-counting.
These four terms
together with the first part of the equation mean that the rate
equation is a non-linear and non-local integro partial-differential
equation:  hence difficult to solve.  In fact, it is interesting to
note that in existing discussions of loops (see for example
\dc{ViSh}),
the effects of terms 1 to 4 are never included ---
that is, the interactions between loops are
ignored (though this is not a very realistic approximation).  The resulting
rate equation is then easy to solve, as we shall see in section
\dr{sscondbeta}.

In a {\em non-expanding universe} we set $\dl = H = 0$ so that
${\paa n}/{\paa t}$ is just given by the four scattering terms
(\dr{full1})-(\dr{fulleqn}).
The rate equation therefore loses its partial
differential aspect.
An important check on this equation
is to verify that the total length
density $L$ in loops defined by
\be
L:=\int_0^{\infty} \nlt \k d\k
\ee
is preserved.\footnote{Length density $L$ and energy density
$E$ are related by $E=\mu L$, where $\mu$ is the energy per
unit length of the string.  In the following we often use the two words
interchangeably.} That is, in a non-expanding universe one should have
\be
\frac{dL}{dt}  =  \int_0^{\infty} \k \frac{\paa \nlt}{\paa t} d\k = 0.
\dle{fundconL}
\ee
By definition each of the $A$ and $B$ scattering processes
individually preserve length (section \dr{bel}).  However,
starting from the rate equation (\dr{fulleqn}) in a non-expanding universe,
it is in fact difficult to prove
(\dr{fundconL}) as one often has to deal with diverging integrals.
For example, if the distribution $\nlt \sim \k^{-5/2}$ for large $\k$,
then the contribution to (\dr{fundconL}) from each of terms 1 to 4 is
divergent.  These need to be carefully regularised in order to
proceed, but once this is done a long
calculation shows that (\dr{fundconL}) is indeed satisfied as required
\dc{Thesis}.

We intend to solve for $\paa n/
\paa t$.
First, however,
the functions $A$,
$B$ and $\dl$ must be estimated.

\section{Determination of
$A$, $B$ and $\dl$}\dle{approx}

We begin by summarising our assumptions about the length scales
expected to characterise the loops.

\subsection{Length Scales}\dle{length}

Throughout this work we assume that there are two length scales on the
loops (see \dc{ACK}).  The first is the persistence length, $\bxi(t)$,
the distance
over which the loops are correlated in direction.  Both in
non-expanding as well as expanding universes, we assume that the loops
are Brownian random walks on scales much greater than
$\bar{\xi}$.  The second scale is denoted by
$\zeta(t)$ and it measures the `wiggliness' of the string
due to the kinks produced as a result of intercommutation
\dc{ViSh,ACK}.

As yet we have not been able to obtain equations for $\bar{\xi}(t)$
and $\zeta (t)$.  The following assumptions are therefore made
about their behaviour:
\begin{itemize}
\item In a {\em non-expanding universe} the two
scales are taken to be time independent; $\bxi = \bxic$ and $\zeta =
\zetac$ where $\bxic$ and $\zetac$ are constants.
\item  In an {\em  expanding
universe} they are assumed to `scale':
\be
\bxi(t) =
\bxic t, \; \; \; \; \zeta(t) = \zetac t .
\ee
Typically, we expect
$\bxi$ to be of the order of the horizon size whilst the smaller
scale $\zeta$ should be characterised by the scale of
gravitational back reaction.  Hence for GUT strings we expect
$\zeta
\simeq \GGmu \bxi \simeq 10^{-5} $ \dc{ACK}.
\end{itemize}

Expressions for
$A$, $B$ and $\dl$ may now be found in terms of $\bar{\xi}$, $\zeta$
and $\k$.  Our aim will then be to try to solve (\dr{fulleqn}) for
$\nlt$ and determine whether, in an expanding universe, the energy
density in loops also scales.  In a non-expanding universe we should
obtain results consistent with string statistical mechanics.

\subsection{Coefficient $A$}

By definition, $A(\k_1+\k_2;\k_1,\k_2)$ is the probability that a
loop of length $\k_1$ intersects a loop of length $\k_2$ to form a
loop of length $\k_1+\k_2$ (see equation (\dr{AAdef})).  We therefore
expect $A$ to take the form
\be
A(\k_1+\k_2;\k_1,\k_2) = \chi \k_1 \k_2
\dle{Acoef},
\ee
where $\chi$ is a relative velocity times a geometric factor --- see
\dc{ACK}.

For a motivation of this
expression, consider two cylinders of lengths $r_1$ and $r_2$ much
greater than their diameter $d$ (so that terms of order $d/r_1$ and
$d/r_2$ can be neglected).  Then, by averaging over all possible
orientations between the cylinders, one can show that their scattering
cross section is proportional to $r_1 r_2$ times their relative
velocity.  If we model the loops as being made up of $\k/
\bxi$ negligibly thin cylinders of length $\bxi$, then the
scattering cross section for two loops of lengths $\k_1$ and $\k_2$
must be of the form given in (\dr{Acoef}).  Note also that
(\dr{Acoef}) has the correct dimensions from
(\dr{loopdim3}).  As in \dc{ACK}, we expect $\chi \sim 0.1 - 0.2$.

\subsection{Coefficient $B$}

This coefficient is much harder to estimate.
However, consider a loop of length
$\k$ and ask for the probability per unit time per unit volume of it
decaying to give a loop of length $y$ with $y<\k/2$.  (From
(\dr{Bsymm}), we then know $B$ for all $y$.)

Firstly suppose that $\bxi \ll y  \ll  \k/2$.
On these scales the loop is, by assumption, a Brownian random walk,
and we take
$B$ to be given by
\be
B(\k-y,y;\k) = \frac{\tilde{\chi} \k}{ (\bar{\xi} y)^{3/2}} \; \; \;
\; \left( \frac{\k}{2} \gg y \gg \bxi \right)
\dle{Bcoef1},
\ee
where $\tilde{\chi}$ is another relative velocity.
To obtain equation (\dr{Bcoef1}), consider
one of the
$\bxi$ segments of the loop, and another such segment $m = y/
\bxi$ steps away.  Then the straight line distance between these
two segments is $m^{1/2}
\bxi$ (assuming Brownian random walks), and the volume contained
between them is of order $(m^{1/2}\bxi)^{3}$.  Thus the
probability that these two segments intersect in time $dt$ to form a
loop of length $y = m  \bar{\xi}$ is equal to the ratio of the volume
swept out by one of them divided by the total volume:
\be
\frac{\tilde{\chi}\bxi^2 dt }{(m^{1/2}\bxi)^{3} }
\dle{Bstep1}.
\ee
Here $\tchi$ is the relative velocity of the two segments.
Since there are $\k/ \bar{\xi}$ pairs of such segments, one obtains
(\dr{Bcoef1}).
Again as in \dc{ACK}, we might expect $\tchi \sim 0.1 - 0.2$
but it is not necessarily equal to $\chi$ as it is a different
relative velocity.

The case for which $y \ll \bxi, \k$ is rather more difficult to
estimate.
Numerical
simulations \dc{AT,BB} suggest that intercommutations lead
to a rapid build up of kinks on these small scales, and therefore
$B(\k-y,y;\k)$ might be expected to peak about $y \sim \zeta$.
(However, if particle production is the dominant mechanism by
which the loops radiate energy --- as has been claimed in 
\dc{MarkGray,MGM} ---
then there will no longer be any small scale structure.  This
interesting case is
not considered here.)    The $B$ coefficient should also be
proportional to the length $\k/\bar{\xi}$ of the loop
and have dimensions given by
(\dr{loopdim4}).  We therefore write
\be
B(\k - y,y;\k) = \frac{1}{\bxi^2} \left( \frac{\k}{\bxi} \right)
h\left( y, \zeta, \bxi  \right)  \; \; \; (y \ll \bxi).
\dle{Bsmall}
\ee
Here $h\left(  y, \zeta, \bxi  \right)$ is a yet unspecified
dimensionless function which is probably peaked about $y \sim \zeta$ and
satisfies $h(0,\zeta,\bxi) = 0$.

Given these forms of $A$ and $B$, we can use the principle of detailed
balance and observe that if an equilibrium situation is reached
in which the rate at which loops break into two daughter loops, $\k +
\k'
\rightarrow \k, \k'$, is equal to the rate at which they join via the
$A$ process, $\k , \k' \rightarrow \k + \k'$, then the corresponding
distribution $n_0(\k)$ must satisfy
\be
A(\k + \k';\k,\k')  n_0(\k)  n_0(\k') = B(\k,\k';\k+\k')  n_0(\k+
\k').
\dle{detbal}
\ee
For large $\k$ equations (\dr{Bcoef1}) and (\dr{Acoef}) show that 
(\dr{detbal})
is satisfied by distributions $n_0(\k) \propto e^{-\beta \k}
\k^{-5/2}$.

In this paper we focus on determining the large $\k$
asymptotic behaviour of solutions to the rate equation,
and it appears that the trends reported in sections
\dr{powerexp} -- \dr{inf} are
in fact not sensitive to the form of $h$ (see \dc{Thesis}).  For
this reason, we simplify calculations
by working with the following
function $B(y,\k-y;\k)$ which we assume to be valid form all $\k,y$:
\be
B(y,\k-y;\k) = \frac{\tchi y (\k-y) (\k+\eta)^{5/2}}{\bxi^{3/2}
(\k-y+\eta)^{5/2}(y+\eta)^{5/2}}.
\dle{Bnew}
\ee
Here $\eta$ is an undetermined length.
First note that by definition $B$ is symmetric as required
(\dr{Bsymm}) and that it has the correct dimensions (\dr{loopdim4}).
Secondly, for $\k \gg y \gg \eta$ (\dr{Bnew}) correctly reduces to
(\dr{Bcoef1}), whereas for $y \simeq \eta \ll \k$ it takes the form
(\dr{Bsmall}) with $h = y \bxi^{3/2} (y+\eta)^{-5/2}$.
Given that (\dr{Bnew}) is also peaked about $y = \k-y \simeq \eta$, we
might
expect $\eta \simeq \zeta$, but we shall leave $\eta$ arbitrary.
However, as with the other lengths characterising the loops, we
assume that $\eta$ scales in an expanding universe with $\eta = a t$
(where $a$ is a constant which should not be confused with the $a$ 
coefficient introduced in section \dr{bel}) and $\eta$ is a constant in a
non-expanding universe.  Observe that, given (\dr{Bnew}), the
corresponding equilibrium distribution
$n_0(\k)$ is
\be
n_0(\k) = \frac{\tchi}{\chi \bxi^{3/2}} \frac{e^{-\beta \k}}{(\k +
\eta)^{5/2}}
\dle{equN}.
\ee

\subsection{Estimating $\dl$}

Expansion of the
universe, redshift and gravitational radiation (G.R) change the length
of a loop.  We
do not consider the loops in the friction dominated era of the
universe though if we did, an extra term would have to be
included to $\dl$ \dc{HiKi,ViSh,CJAP}.  Also, the possibility that loops
may decay predominantly by particle production is not
taken into account \dc{MarkGray,MGM}.

The
contribution to $\dl$ from redshift and expansion is \dc{ViSh}
\be
\left. \dl \right|_{expansion} =  H \left( 1 - 2
\langle v^2 \rangle_{\k,\zeta} \right) \k
\dle{ldotexp}.
\ee
Here $ \langle v^2 \rangle_{\k,\zeta}$ is the
average velocity squared on the loop.  As a first
approximation,
we take
$
\langle v^2
\rangle_{l,\zeta}$ to be a constant with $1 - 2 \langle v^2
\rangle_{l,\zeta} = \alpha$.  From \dc{BB} we might expect that
$\alpha \sim 0.1$ in the radiation era, and $\alpha \sim
0.26$ in the matter eras.\footnote{However, since there are no infinite
strings here, we expect $\alpha$ actually to be rather smaller than
these values which will then represent upper bounds.}
The contribution from gravitational radiation is modelled by
\be
\left. \dl \right|_{G.R.} =  -\Gamma G \mu f\left(
\frac{\k}{\zeta}\right)
\dle{ldotgr}.
\ee
Here $\Gamma$ is a constant, and $f\left(
\frac{\k}{\zeta}\right)$ is a function which tends to 1 as $\k$ tends
to zero, and for large $\k$ is proportional to $\k/\zeta$.  The reason
for this latter limit is that the dominant source of
gravitational radiation from long loops is from the kinks \dc{HGr}, and
the ratio $\k/\zeta$ just counts the number of kinks on a loop of
length
$\k$.  The former limit comes from applying the quadrupole formula to
a loop in Minkowski space \dc{HiKi}.  The simplest form of
$f$, the one we assume, is
$f\left(
\frac{\k}{\zeta}\right) = 1 + m\frac{\k}{\zeta}$ where $m$ is a
constant.  Of course it will be
important to check the sensitivity of solutions of the rate
equation to this choice of $f$, and also to the approximation made for
$B$.

Summarising, we write
\be
\dl = - \Gamma G \mu + \frac{K\k}{t},
\dle{ldotfull}
\ee
where
\be
K = -\frac{m \Gamma G \mu
}{\zetac} + \frac{\alpha}{2} .
\dle{keq}
\ee
Given the values of $\alpha$, we expect $K \, \lsim \,
0.1$.  For long (many times the horizon size) loops, we also expect the
dominant length changing effect to be from the expansion of the universe,
and therefore that $K>0$.  Another reason for taking $K>0$ is given in
the next section.

In a
non-expanding universe, we set $\GGmu = K = 0$ so that $\dl = 0$.

Given these estimates for the coefficients $A$, $B$ and $\dl$, the
rate equation may now be analysed.  However, before doing this a final
change of variables is carried out so as to clarify the problem in an
expanding universe.

\section{Rate equation in scaling variables}\dle{sscalvar}

In the previous section we have made the assumption that in an
expanding universe, the two scales
$\bxi$ and
$\zeta$ are proportional to $t$.  Since we aim to see whether the
energy density in loops also scales in that case ({\em i.e.}\ $L
\propto t^{-2}$), it is useful in an expanding universe\footnote{This
change of variables does {\em not} make sense in a non-expanding universe.
There one must continue to work with (\dr{fulleqn}) in which $\dl = H
= 0$.}
to change variables from
$\k$ and $\nlt$ to `scaling' variables, $z$ and $\gzt$ defined by
\be
z = \frac{\k}{t}
\dle{zdef}
\ee
and
\be
 \gzt = {t^4} \nlt.
\dle{gdef}
\ee
Notice that $z$ is dimensionless, and for
lengths which `scale' it is a constant.   Also observe that
\be
L(t) = \int_0^{\infty} \k \nlt \dk = \frac{1}{t^2} \int_0^{\infty} z
\gzt dz =: \frac{E_{sc}(t)}{\mu t^2},
\dle{Edef}
\ee
where $E_{sc}$ is the energy density in scaling variables.  Thus the
energy density in loops scales when $\gzt = g(z)$;  these are the
scaling solutions we are searching for.  Now, using the definition of the
coefficients $A$ and $B$, one has that
\ba
A(zt;z't,(z-z')t) &= &t^2 \Ai,
\dla{Ascaling}
\\
B(zt,(z'-z)t;z't;\bxic t) & = & \frac{1}{t^2} B(z,z'-z;z';\bxic)
\dla{Bscaling},
\ea
where for the moment we have exhibited $\bxi$ as a fourth argument.
Thus a rate equation for $\gzt$ can be obtained by
substituting into (\dr{fulleqn}) and using
equation (\dr{ldotfull}).
It is
\ba
t \frac{\paa \gzt}{\paa t}& = &
\gzt \left( 4 - \frac{3}{p} - K \right) +
\frac{\partial \gzt}{\partial z} \left( z(1-K) + \GGmu \right)
\dle{ga}
\\
& - &
\GGmu \delta(z) [\gzt]|_{z \rightarrow 0^+}
\dle{gb}
\\
 & +
& \int_{0}^{z/2} \Ai \gzpt g(z-z',t)dz'
\dle{g1}
 \\
& + &  \int_{z}^{\infty} \Bii \gzpt dz'
\dle{g2}
 \\
& - & \gzt  \int_{0}^{\infty} \Aiii \gzpt dz'
\dle{g3}
 \\
& - & \gzt  \int_{0}^{z /2}  \Biiii dz'
\dle{geqn}.
\ea
Here the parameter $p$ is defined by
the expansion rate; $R \propto t^{1/p}$, so that $p=2$ in the
radiation  era, and $p=\frac{3}{2}$ in the matter era.
Observe that the RHS of the above equation is not explicitly $t$
dependent and hence that it is consistent to have
$\gzt = g(z)$ -- {\em i.e.}\ a scaling solution is consistent
with the equation.  It remains to find the form of this scaling
solution and to see whether it is stable.  Henceforth we always work
with the scaling variables $z$ and $\gzt$ in an expanding universe.

An equation for $dE_{sc}/dt$ can be obtained by using equations
(\dr{Edef}) and (\dr{fundconL}), and integrating
(\dr{geqn}).  It is
\be
\frac{dE_{sc}}{dt} = \left( 2 + K - \frac{3}{p} \right)\frac{E_{sc}}{t} -
\frac{\Gamma G \mu^2  N_{sc}}{t}.
\dle{Escaling}
\ee
Here $N_{sc}(t)=t^3 N(t)$ where $N(t)=\int_0^{\infty} \nlt d\k$
is the total number of loops
at time $t$.  Thus it is possible for the energy density
to scale both in the radiation and matter eras provided $K>0$, and
$\GGmu \neq 0$.  We therefore take
$K > 0$, whilst the important r\^{o}le played by $\GGmu$ will again
become apparent in section \dr{powerexp}.
Note that
the assumption of scaling lengths $\bxi$ and $\zeta$ does {\em
not} automatically imply that the energy density in loops also
scales in an expanding universe:
this will only happen if values of $E_{sc}$ and $N_{sc}$ are
obtained such that the right hand side of (\dr{Escaling}) vanishes.
The aim of the rest of this paper is to see whether such scaling
solutions exist and whether they are stable.

The work
presented here is entirely analytical and given the complicated
nature of the rate equation we do not attempt to find any exact
solutions but instead focus on their large $z \gg \bxi$
asymptotic behaviour.
We are currently
trying to develop methods to solve (\dr{geqn}) numerically \dc{Joao}.

\section{Search for solutions $\gzt = z^{\alpha} e^{-\beta
z}$}\dle{powerexp}

In the rest of this paper we focus on the functional form of large
$z$ solutions of the rate equation.  This section in particular tries
to determine whether stable scaling solutions of the form
\be
\gzt = {c(t) e^{-\beta(t) z}}{z^{\alpha(t)}} \; \; \; \; \beta > 0,
\dle{hope}
\ee
exist for large $z \gg \bxic$.

The motivation for
looking for such solutions comes from results of string statistical
mechanics \dc{MT,Edandco,SakVil}:  in a non-expanding universe
the loop network evolves
until it reaches a stable equilibrium distribution, and
if the energy density in loops is less than a critical value, this
distribution is given by\footnote{Recall that in a non-expanding
universe $z=\k$ and $\gzt = \nlt$.}
\be
g(z) = \frac{c e^{-\beta z}}{z^{5/2}}
\dle{equilstatmech}
\ee
for large $z$ \dc{MT,Edandco}.  This form cannot of course be correct for
$z \rightarrow 0$ since it would lead to infinite energy.  Hence
it must be modified for small $z$.  In (\dr{equilstatmech}) $\beta \geq 0$
since the energy in loops must be
finite.  At $\beta = 0$ the energy density in loops is equal to the
critical value and the loops have a scale invariant distribution $\gzt
\propto z^{-5/2}$.  A phase transition then occurs, and the system is
unstable against the formation of an infinite string
\dc{MT,Edandco,SakVil}.

Since our model of loop evolution can be applied to
loops in a non-expanding universe, it should also predict stable
solutions of the form (\dr{equilstatmech})
as long as the energy
density in loops is below the critical value
(which in turn should be determined
by the parameters in the model).  In this section we therefore use these
comments as motivation to look for large
$z$ solutions of the rate equation of the form (\dr{hope})
{\em both} in a
non-expanding as well as an expanding universe.
Here we only consider the
case for which $\beta > 0$.  Critical power law loop distributions
with $\beta = 0$ are analysed in the next section.

This section contains a number of parts.  In the first we try to get a
feel for the rate equation by
considering a simplified, but unphysical, limit of this equation.
In this case we show that the equation admits solutions of the form
(\dr{hope}) for large $z$.  The conditions required
for these solutions to be scaling solutions, $\gzt = g(z)$, are also
specified, and comments are made about stability.  The aim of section
\dr{full} is
to determine whether a distribution of the
form (\dr{hope}) is consistent with the full rate equation.
To do that, the RHS of the rate equation is calculated for
distributions (\dr{hope}), and consistency conditions allow equations
for $\dot{\alpha}$, $\dot{\beta}$ and $\dot{c}$ to be obtained.  These
are then analysed in sections
\dr{snonexpbeta}-\dr{sexpbeta}.  In section
\dr{snonexpbeta} we work in a non-expanding universe.  We see that as
long as the energy density in loops is below a critical value,
$E_{crit}$ --- which is expressed in terms of the parameters of the
model --- there is a stable scaling solution of the form
(\dr{equilstatmech}).  The corresponding value of
$c$ is determined, and we show
that if the energy density is greater that $E_{crit}$, there is no
stable solution (\dr{equilstatmech}).  The results of statistical
mechanics suggest that infinite strings may be formed in that
case.  An expanding universe is considered in section
\dr{sexpbeta} and we investigate whether a scaling solution of the
form (\dr{hope}) can
exist in the radiation and matter eras.

\subsection{Conditions for a solution $\gzt = {c(t) e^{-\beta(t)
z}}{z^{\alpha(t)}}$}\dle{sscondbeta}

We begin the search for large $z$ solutions of the rate equation
(\dr{geqn}) by considering a simple, unphysical, limit of that
equation in which there are no interactions between the
loops; $A=B=0$.  In the radiation era $p=2$ the rate
equation (\dr{geqn}) now reduces to
\be
t \frac{\paa \gzt}{\paa t}  =  \gzt \left( \frac{5}{2} - K \right)
+ (1-K) z \frac{\paa \gzt}{\paa z}
+ \GGmu \frac{\paa \gzt}{\paa z} -
\GGmu \delta(z) [\gzt]|_{z \rightarrow 0^+}
\dle{eqGgmub}.
\ee
Without solving (\dr{eqGgmub}) explicitly, one can investigate whether
it has solutions of the form (\dr{hope}) for large $z$ and all times
by simply substituting the ansatz $\gzt =c(t) e^{-\beta(t)
z}{z^{\alpha(t)}} $ into (\dr{eqGgmub}).  For $z>0$ it gives
\be
t \frac{\dot{c}}{c}  - t z\dot{\beta} + t \dot{\alpha}  \ln z  =
\left( \frac{5}{2} - K + \alpha(1-K) - \GGmu \beta \right) -
z\beta(1-K)    +
\frac{\alpha \GGmu}{z}.
\dle{hope3}
\ee
In the large $z$ limit where
$ \left| {\GGmu \alpha(t) }/{ z}\right| $ is negligible compared
to all the other terms, one can equate the coefficients of
$z^0$, $z$ and $\ln z$ in (\dr{hope3}) to obtain equations
for $\dot{\alpha}$,
$\dot{\beta}$ and $\dot{c}$.  They are
\ba
\dot{\alpha} & = & 0,
\dla{alphaeq}
\\
t \dot{\beta} &=& (1-K) \beta \; \; \Longleftrightarrow \beta(t) =
{\beta(t_0)}\left(\frac{ t}{t_0} \right)^{1-K},
\dla{betaevol}
\\
t \frac{\dot{c}}{c} & = & \frac{5}{2} - K  + (1-K) \alpha - \beta
\GGmu.
\dla{cevol}
\ea
Generally in this section we require $\beta > 0$.  However, in this
special case we may note that if $\beta(t_0) = 0$ then $\beta$ stays zero
for all times.  In that case a scaling solution ($\dot{\alpha} =
\dot{\beta} = \dot{c} =0$) can only exist for the special value
$\alpha =
-( 5/2 - K)/(1-K)$.  Such power law distributions are discussed
further in section
\dr{inf}.

If
$\beta(t_0) > 0$ then $\beta$ grows with time (as $1-K>0$) and so the
system never reaches a scaling solution.  From
equation (\dr{cevol}) observe that when
$\beta \simeq 1/\GGmu$, the dominant term on the RHS of
that equation is given by $\beta \GGmu$ , and so $c
\rightarrow 0$ as $t \rightarrow \infty$.  Thus $\gzt \rightarrow 0$
as $t \rightarrow \infty$ and if this were a realistic model of
loop evolution, we would say that all the loops disappear.  The
solution $\beta = 0$ is therefore an unstable one.

This analysis is confirmed by calculating the exact solution to
(\dr{eqGgmub}).  For an initial distribution at $t=t_0$ given by
$g(z,t_0) =
c(t_0) e^{-\beta_0 z}z^{-5/2}$, it is
\be
\gzt = c(t_0) \left(\frac{t}{t_0} \right)^{1-K}
\frac{e^{-\beta_0 \GGmu \left(\frac{t}{t_0} - 1 \right)} e^{-\beta_0
z}}{ \left[ z + \GGmu \left( \frac{t}{t_0} - 1 \right) \right]^{5/2} }
\theta (z).
\dle{exactsoln}
\ee
We comment that the effect of the term involving a $\delta$ function
in (\dr{eqGgmub}) is the appearance of the $\theta$ function in the
solution (\dr{exactsoln}).  Hence the $\delta$ function
ensures that $\gzt = 0$ $\forall
z<0, \forall t$ as was claimed in section \dr{bel}:  had the
$\delta$ function not been there, then there would be no $\theta$ function
in (\dr{exactsoln}) and hence the solution $\gzt$ would have been
unphysical, since
for any $t>t_0$, $\gzt \neq 0$ when $-\GGmu(t/t_0 -1) < z < 0$.

We stress that as long as $\beta \neq
0$, the factor $\GGmu$ plays a very important r\^{o}le in
determining the large $t$ behaviour of $c(t)$.
This is consistent with the
comments of section \dr{sscalvar}.

Finally it is interesting to note that if there was an
additional term $P \gzt$ on the RHS of (\dr{eqGgmub})
that did not vanish for $\beta = 0$, then $\beta = 0$ would no
longer be a solution to the equation.  For negative $P$ the solution
would instead be an unstable one with $\beta \neq 0$.

\subsection{The full rate equation}
\dle{full}

Now consider the effect of adding the four scattering terms to the
RHS of (\dr{eqGgmub}).  However, rather than working with a single
value of $\alpha$ as in (\dr{hope}), we work with a more general
distribution function
\be
\gzt = \theta (z) \frac{e^{-\beta(t) z}}{(z+a)^{5/2}} \sum_{n=0}^{\infty}
\frac{c_n(t)}{(z+a)^n} .
\dle{hope4}
\ee
This can be regarded as a modification of (\dr{equN}).
We wish to see if such distributions
are consistent with the rate equation, and if so to obtain
equations for the $\dot{c}_n$ and $\dot{\beta}$.  Note that if there
is a solution $\gzt = {c_0 e^{-\beta z}}{(z+a)^{-5/2}}$ then we expect
the $c_n \rightarrow 0$ ($n \geq 1$) as $t \rightarrow \infty$ whilst
$c_0(t)$ tends to a constant.

Equations for
$\dot{\beta}$ and $\dot{c}_n$ are obtained by substituting the ansatz
(\dr{hope4}) into the full rate equation, expanding in powers of
$\frac{a}{z}$, and then equating the coefficients of $z$, $z^0$,
$z^{-1}$ $\ldots$.  As an example of how to deal with the scattering
integrals, consider term 2 which is given
by
\ba
{\rm Term \; 2} &=&  \frac{\tchi z}{\bxic^{3/2} (z+a)^{5/2}}
e^{-\beta z} \sum_{n=0}^{\infty} c_n  \int_{0}^{\infty}
\frac{y  e^{-\beta y} }{ (y+z+a)^{n} (y+a)^{5/2}} dy .
\dla{term2newB}
\ea
We are interested in the $z \gg a$ limit as we are searching for large
$z$ solutions to the rate equation.
If $\beta > 0$ (the case
considered in this section), the exponential inside the integrand
naturally provides a  cutoff at $ x \sim 1/\beta$, so that provided we
look at large $z \gg 1/\beta$, the denominator of
(\dr{term2newB}) can be Taylor expanded in powers of $\frac{y}{z}$.
The result is
\ba
{\rm Term \; 2} &=&  \frac{\tchi}{\bxic^{3/2}} e^{-\beta z}
 \frac{1}{z^{3/2}}  \sum_{n=0}^{\infty}
\frac{c_n}{z^n} \left[
\bullet + \bullet
\frac{1}{z} + \bullet \frac{1}{z^2} + \ldots \right] \; \; \; \;
\left(z
\gg a, \frac{1}{\beta} \right) ,
\dla{expanaly}
\ea
where each $\bullet$ is an $n$ dependent coefficient arising from the
Taylor expansion.
The case for which $\beta = 0$ and $n\neq 0$ (and hence
such a Taylor expansion is not
permissible) is considered in section \dr{inf}.

After a similar analysis for the other scattering integrals, equations
can be obtained for $\dot{\beta}$ and $\dot{c}_n$.
To zeroth order, we set $c_n = 0$ for $n \geq 1$ and obtain
equations for $\beta$ and $c_0$ from the coefficients $z$ and $z^0$
respectively in the rate equation (\dr{geqn}).  They are\footnote{We
set $\bxic = 1$ in the rest of this paper so as not to have to write
explicitly factors of $\bxic$.  These can be reinstated into the
following equations by changing $\tchi
\rightarrow \tchi/\bxic^{3/2}$.}
\ba
t\dot{\beta} & = & (1 - K) \beta + \left( \chi c_0 - \tchi
\right) \left[I_1  -  \frac{4}{ 3 a^{1/2}} \right],
\dla{betaeq}
\\
t\frac{\dot{c_0}}{c_0} &=&
\left(\frac{3}{2} - \frac{3}{p} + \frac{3K}{2} \right) - \GGmu
\beta - 8 \sqrt{a} (\chi c_0 - \tchi),
\dla{czeroeq}
\ea
where
\be
I_1 = \int_0^{\infty} \frac{y e^{-\beta y}}{(y+a)^{5/2}} dy =
e^{\beta a} \sqrt{\beta} \Gamma(-1/2,\beta a) - a e^{\beta
a} \beta^{3/2} \Gamma(-3/2,\beta a).
\ee

To first order (where $c_n = 0, n\geq 2$) we have a more
complicated set of equations valid for $\beta > 0$;
\ba
t \dot{\beta} & = & (1-K)\beta + (\chi c_0 - \tchi) \left[ I_1 -
\frac{4}{3a^{1/2}} \right] + \chi c_1 \left[ {I}_2 - \frac{4}{15
a^{3/2}} \right],
\dla{bcompeqn}
\\
t \dot{c}_0 & = & \left( \frac{3}{2} - \frac{3}{p} + \frac{3K}{2}
\right)c_0 - \GGmu \beta c_0 - 8 c_0 a^{1/2} (\chi c_0 - \tchi)
+ \frac{8}{5} \frac{\chi c_0 c_1}{a^{1/2}},
\dla{ccompeqn}
\\
t \dot{c}_1 & = & \left( \frac{1}{2} - \frac{3}{p} + \frac{5K}{2}
\right) c_1 - \GGmu \left( \frac{5c_0}{2} + \beta c_1 \right)
 + \frac{100}{3} a^{3/2} c_0 (\chi c_0 - \tchi)
\nonumber
\\
&&+ \chi c_1 \left(
\frac{8c_1}{3a^{1/2}} - 28 c_0 a^{1/2} \right)
+ \tchi(8a^{1/2}c_1 - c_1 I_3) + \frac{5}{2}(1-K)ac_0 .
\dla{c1compeqn}
\ea
Here
\be
{I}_2 = \int_0^{\infty} \frac{y e^{-\beta y}}{(y+a)^{7/2}} dy =
e^{\beta a}\left( 1 + \frac{2a\beta}{5} \right) \beta^{3/2}
\Gamma(-\frac{3}{2}, \beta a) - \frac{2 }{5a^{3/2}},
\ee
and
\be
I_3 = \int_0^{\infty} \frac{y^2 e^{-\beta y}}{(y+a)^{5/2}} dy =
e^{\beta a} \left[ \beta^{-1/2} \Gamma(1/2,\beta a) -
2a\beta^{1/2}\Gamma(-1/2,\beta a) + a^2 \beta^{3/2} \Gamma(-3/2,\beta
a) \right].
\ee

\subsection{Analysis in a non-expanding universe}
\dle{snonexpbeta}

Consider first the behaviour of equations
(\dr{betaeq})-(\dr{c1compeqn})
in a non-expanding universe where we should obtain
results consistent with string statistical mechanics.  For the
purposes of an analytical analysis, we assume that $\beta$ is small.
Then the zeroth order equations (\dr{betaeq}) and (\dr{czeroeq})
reduce to
\ba
\dot{\beta} = - 2 \sqrt{\pi} \sqrt{\beta} (\chi c_0 - \tchi),
\dla{betanonexp}
\\
 \dot{c_0} = - 8 \sqrt{a} (\chi c_0 - \tchi)  c_0.
\dla{cnonexp}
\ea
Observe that a
stationary solution with $\dot{c}_0 = \dot{\beta} = 0$
exists when $c_0 =c_0^{crit}$ where
\be
c_0^{crit} = \tchi /\chi.
\dle{ccritdef}
\ee
As expected, the value of $\beta$ is not fixed by the value of
$c_0$ in this case:  these two variables are linked by the total
energy density of the system which is conserved by the rate equation
(section \dr{deriv}).  Also observe that $c_0 >
\tchi/\chi \Rightarrow \dot{c_0}<0$, $\dot{\beta} < 0$ and similarly
$c_0 < \tchi/\chi \Rightarrow \dot{c_0}>0$, $\dot{\beta} > 0$ so that
this solution is a stable one -- as is required for equilibrium
distributions.

Figure \dr{phasenonexp} shows trajectories of constant energy density
in the
$(c_0,\beta)$ plane.
\typeout{figure: phase diag non-exp}
\begin{figure}[htb]
\setlength{\unitlength}{0.1cm}
\begin{picture}(0,0)
\put(127,-20){\makebox(0,0)[lb]{$1.5 E_{crit}$}}
\put(127,-40){\makebox(0,0)[lb]{$ E_{crit}$}}
\put(127,-60){\makebox(0,0)[lb]{$ 0.5E_{crit}$}}
\put(127,-69){\makebox(0,0)[lb]{$ 0.3E_{crit}$}}
\put(128,-90){\makebox(0,0)[lb]{\large $\beta$}}
\put(28,-16){\makebox(0,0)[lb]{\large $c_0$}}
\end{picture}
{\centerline {\epsffile{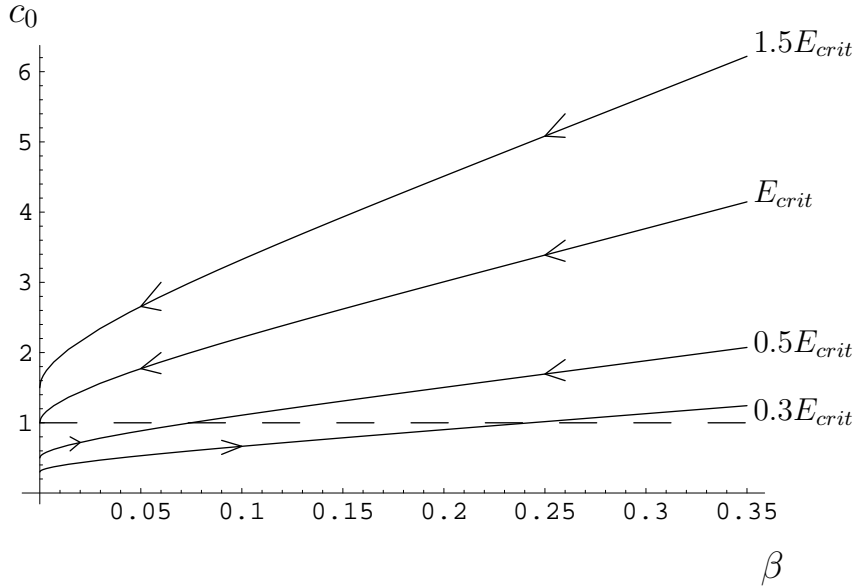}}}
\caption{The $(c_0,\beta)$ phase diagram showing lines of constant
energy (solid curves).  The parameters are $a=1$ and $\chi = \tchi =
1$ so that $c_0^{crit} = 1$.  The line of stable solutions is given
by the dashed curve.}
\label{phasenonexp}
\end{figure}
We observe that as long as the energy density $E$ is less than a
critical value $E_{crit}$ given by (here $c_0 = c_0^{crit}$ and $\beta
= 0$)
\be
E_{crit}  = \int_{0}^{\infty} \frac{\tchi}{\chi}
\frac{z}{(z+a)^{5/2}} dz = \frac{\tchi}{\chi} \frac{4}{3\sqrt{a}},
\ee
then the stable scaling solution is reached since the trajectories of
constant energy cross the stable equilibrium line for which $c_0 =
c_0^{crit}$.
If $E \, \gsim \, E_{crit}$ then $\beta \rightarrow 0$, but at $\beta = 0$
the equations (\dr{betanonexp}) and (\dr{cnonexp}) break down
(see section \dr{inf}).
However,  we recall that from string statistical mechanics \dc{MT} the
point $\beta = 0$ corresponds to a phase transition and
the formation of an infinite string.  If this occurs, we cannot analyse
the evolution of the system further as our equations only model a
network of loops which does not contain infinite strings.  This case
is considered in section \dr{inf}.

Our conclusions on the evolution of $\beta$ and $c_0$ remain
unchanged by the analysis of the more complicated equations
(\dr{bcompeqn})-(\dr{c1compeqn}) containing $c_0$,
$c_1$ and $\beta$.  Indeed, we see that $c_1 \rightarrow 0$ for all
initial values of the parameters whilst $\beta$ and $c_0$ behave as
discussed above (see figures \dr{nonup} and \dr{noncrash}).

\begin{figure}[ht]
\centerline{\hbox{
\psfig{figure=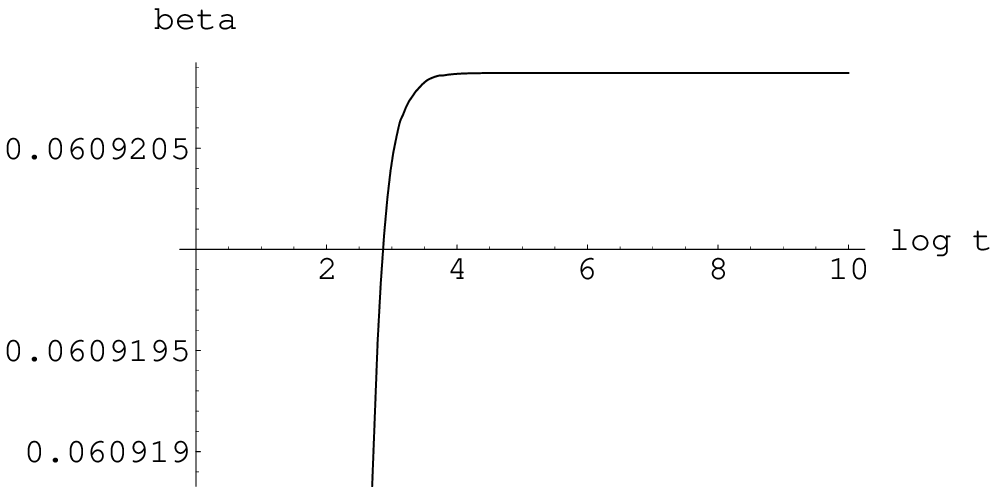,height=5.5cm,width=5.5cm}
\psfig{figure=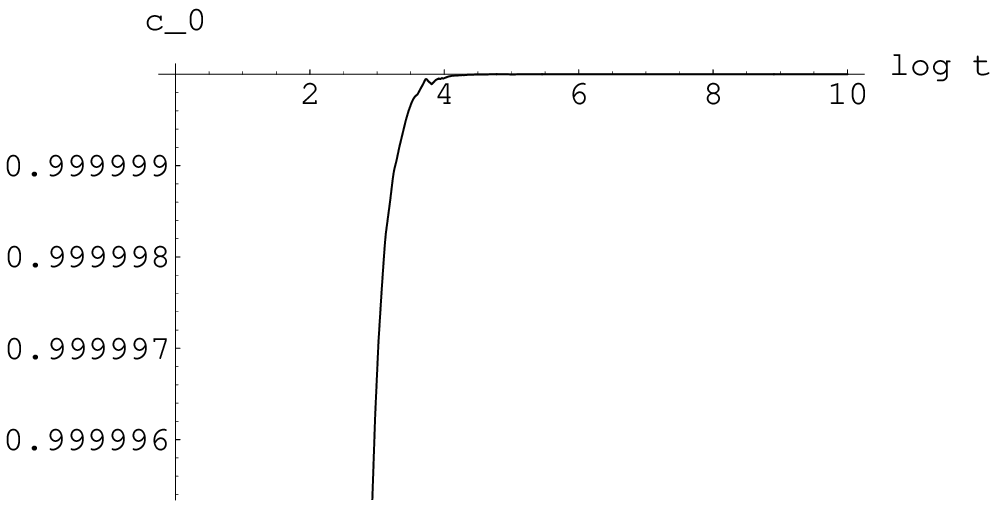,height=5.5cm,width=5.5cm}
\psfig{figure=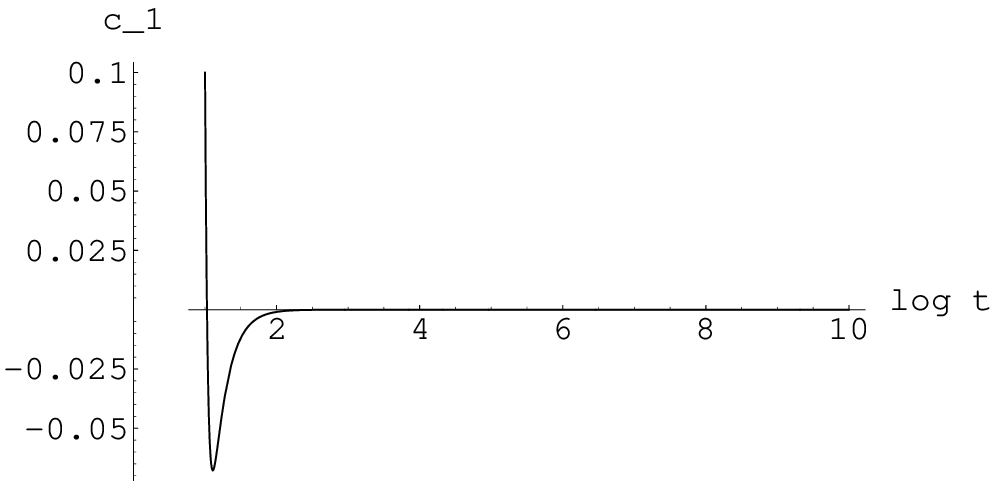,height=5.5cm,width=5.5cm}}}
\caption[Evolution of $\beta$, $c_0$ and $c_1$.]{Evolution 
of $\beta$, $c_0$ and $c_1$ according to equations
(\dr{bcompeqn})-(\dr{c1compeqn}) adapted
for a non-expanding universe.  The parameters are $\tchi = \chi = 1$
so that $c_0^{crit} = 1$, and $a=1$ giving a critical energy of
$E_{crit} = 1.33$.  The initial values of the
variables at time $t=1$ are $\beta(1) = 0.05$, $c_0(1)=0.9$,
$c_1(1)=0.1$ giving a constant energy of $E = 0.887 < E_{crit}$.
Since the initial
value of $c_0$ is less than $c_0^{crit}$, both $c_0$ and $\beta$ are
seen to increase
according to figure (\dr{phasenonexp}) while $c_1 \rightarrow 0$.}
\label{nonup}
\end{figure}
\begin{figure}[ht]
\centerline{\hbox{
\psfig{figure=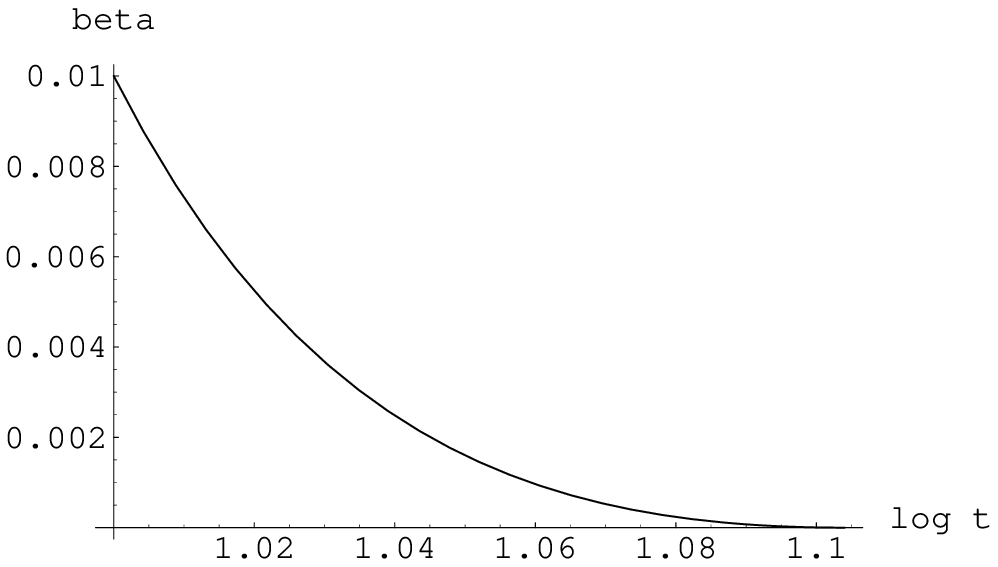,height=5.5cm,width=5.5cm}
\psfig{figure=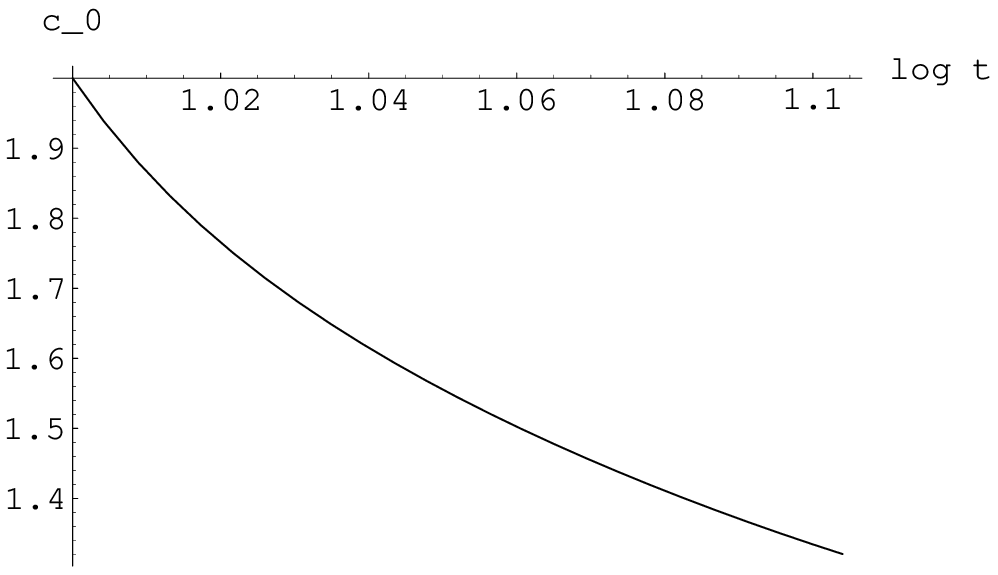,height=5.5cm,width=5.5cm}
\psfig{figure=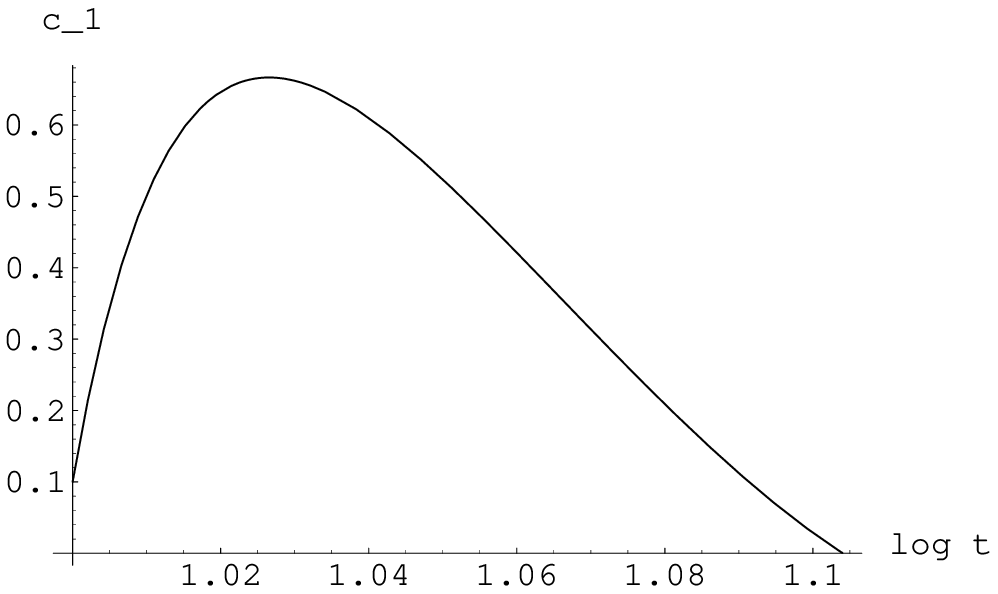,height=5.5cm,width=5.5cm}}}
\caption{This figure is similar to the previous one except that now
$\beta(1) = 0.01$, $c_0(1)=2$, $c_1(1)=0.1$.  Hence $E=2.57 >
E_{crit}$.  As expected, $\beta$ decreases until it reaches zero,
$c_0$ also decreases whilst $c_1$ eventually tends to $0$.
At $\beta = 0$ the equations are no longer valid.}
\label{noncrash}
\end{figure}

To conclude, in a non-expanding universe we obtain results consistent
with those of statistical mechanics {\em provided} the total energy
in the loop network is less than $E_{crit}$.

\subsection{Analysis in an expanding universe}\dle{sexpbeta}

In an expanding universe, the small $\beta$ limit of equations (\dr{betaeq})
and (\dr{czeroeq}) is
\ba
t \dot{\beta} = (1-K)\beta - 2 \sqrt{\pi} \sqrt{\beta} (\chi c_0 -
\tchi),
\dla{betaexp}
\\
t \frac{\dot{c_0}}{c_0} = \left( \frac{3}{2} - \frac{3}{p} +
\frac{3}{2}K
\right) - \GGmu \beta - 8 \sqrt{a} (\chi c_0 - \tchi) .
\dla{cexp}
\ea
The scaling solution is given by $\dot{\beta} = \dot{c_0} = 0$.
In the radiation era these two conditions are compatible
--- we will discuss the
stability properties (or lack thereof) of this solution in the
next subsection.
In the matter era,
however, no full scaling solution is permitted.
The reason for this difference between the two eras can be seen
from (\dr{betaexp})-(\dr{cexp}).  Imposing the conditions
$\dot{c_0} =
\dot{\beta} = 0$ gives
\be
(1-K) \sqrt{\beta}= \frac{3}{4} \sqrt{\frac{\pi}{a}} \left[ \left(
\frac{1}{2} - \frac{1}{p} + \frac{K}{2} \right) - \frac{\GGmu \beta
}{3} \right].
\ee
If $\GGmu\beta$ is very small compared to the other terms (we expect
$\GGmu \simeq 10^{-5}$ and $\beta$ to be initially small), the
condition for  a scaling solution to exist is therefore
\be
\left( \frac{1}{2} - \frac{1}{p} + \frac{K}{2} \right) > 0.
\dle{condscal}
\ee
Thus in the radiation era ($p=2$), a scaling solution can exist as 
$K > 0$.  In
the matter era,
however, condition (\dr{condscal}) requires $K > 1/3 \, \gsim \, 0.33$.
Since we expect $0 < K \, \lsim \, 0.1$, this therefore implies that
{\em no scaling solution is possible in the matter era}.

Equations (\dr{betaexp})-(\dr{cexp}) are now analysed in the
radiation and matter eras.

\subsubsection{Radiation era}

In the radiation era, scaling occurs when
$\beta = \beta_{crit}$ and $c_0 = c_0^{crit}$ where
\ba
(\chi c_0^{crit} - \tchi) &=& \frac{3K}{16 \sqrt{a}},
\dla{ccrit}
\\
\beta_{crit} & = & \frac{9\pi K^2}{64a (1-K)^2}.
\dla{betacrit}
\ea
Here we have ignored terms of order $\GGmu \beta$.  In this limit,
equation (\dr{cexp}) can be written as
\be
t \frac{\dot{c_0}}{c_0} = 8 \chi \sqrt{a} ({c}_0^{crit} -  c_0) .
\dle{star}
\ee
Hence $c_0 =c_0^{crit}$ is an
attractor since $c_0 > c_0^{crit} \Rightarrow \dot{c_0} < 0 $ and $c_0
< c_0^{crit} \Rightarrow \dot{c_0} > 0 $.
However, analysis of equations (\dr{betaexp}) and (\dr{cexp}) shows
that, depending on the initial values of $\beta$ and $c_0$ at time $t=t_I$,
there are two very different
possible scenarios for the evolution.

Consider first the case in which $c_0(t_I) = c_0^{crit}$.
If $\beta(t_I) > \beta_{crit}$ then the dominant term in (\dr{betaexp}) is
the first one, and so $\beta$ increases.  Even though equations
(\dr{betaexp}) and (\dr{cexp}) are only valid
for small $\beta$, the full equations (\dr{betaeq}) and (\dr{czeroeq})
show that $\beta$ keeps on and on growing.  When $\beta \simeq
1/\GGmu$, then the dominant term on the RHS of (\dr{cexp}) is $-\GGmu
\beta$ and so for large times, $c_0 \rightarrow 0$, $\beta \rightarrow
\infty$:  the loops {\em disappear}.
If $\beta < \beta_{crit}$ then we see from (\dr{betaexp}) that $\beta
\rightarrow 0$ whilst $c_0 \rightarrow c_0^{crit}$.
The scaling
solution $\beta = \beta_{crit}$ and $c_0 = c_0^{crit}$ is therefore an
unstable one.
For arbitrary values of $c_0(t_I)$ and $\beta(t_I)$, the situation is
summarised in figure \dr{rad}.
\begin{figure}[htb]
\setlength{\unitlength}{0.1cm}
\begin{picture}(0,0)
\put(49,-63){\circle*{1}}
\put(110,-50){\makebox(0,0)[lb]{\large $\beta \rightarrow \infty$}}
\put(110,-55){\makebox(0,0)[lb]{\large $c_0 \rightarrow 0 $}}
\put(55,-30){\makebox(0,0)[lb]{\large $\beta \rightarrow 0$}}
\put(55,-35){\makebox(0,0)[lb]{\large $ c_0 \rightarrow c_0^{crit}$}}
\put(132,-75){\makebox(0,0)[lb]{\large $\beta(t_{I})$}}
\put(30,-16){\makebox(0,0)[lb]{\large $c_0(t_I)$}}
\end{picture}
{\centerline {\epsffile{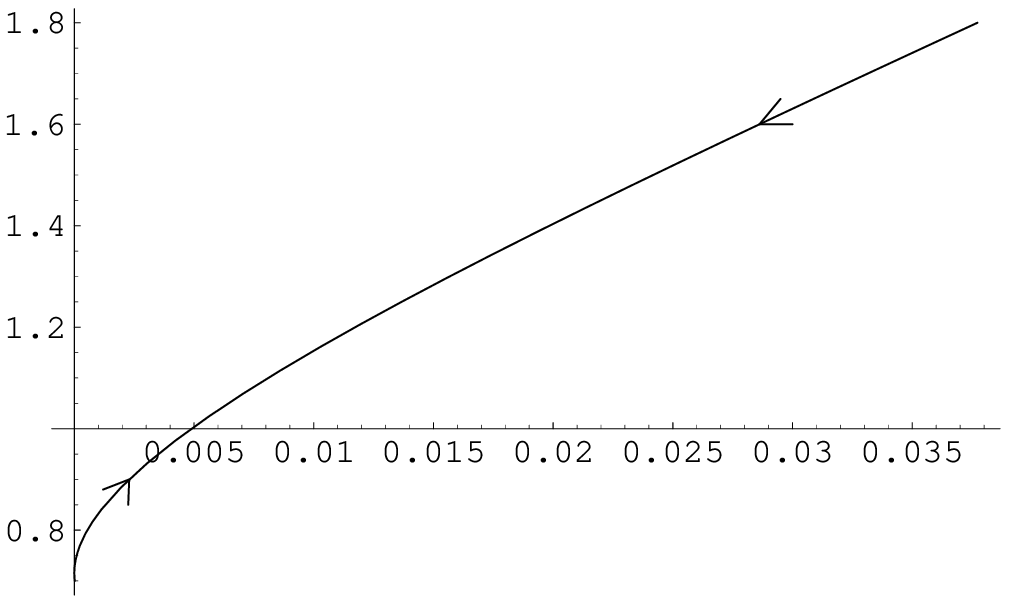}}}
\caption{The $(c_0,\beta)$ plane at the initial time $t=t_{I}$
in the radiation era.  The parameters are $\chi = \tchi = 1$, and
$a=1$ so that the critical point, marked with a dot in the figure, is at
$c_0^{crit} = 1.0187$ and $\beta_{crit} = 0.00545$.  A scaling
solution with $\beta = \beta_{crit}$ and $c_0 = c_0^{crit}$ is reached
for initial
values of $c_0(t_I)$ and $\beta(t_I)$ lying along the curve.  However,
this curve is unstable, so that for initial values deviating slightly
from these $\beta \rightarrow \infty; c_0 \rightarrow 0$ or
$\beta \rightarrow 0; c_0 \rightarrow c_0^{crit}$ as indicated.
The effect of decreasing $a$ is to increase the region of
parameter space for which $\beta \rightarrow 0$.
}
\label{rad}
\end{figure}
Comparison of figures \dr{rad} and \dr{phasenonexp} shows that in both
cases, $\beta \rightarrow 0$ for high loop densities.  However, for low
densities the loops reach an equilibrium configuration in a
non-expanding universe but disappear in a radiation
dominated universe.

Analysis of equations (\dr{bcompeqn})-(\dr{c1compeqn}) shows that the
effect of adding the $c_1$ contribution is to
change slightly the
position of the line of scaling solutions in figure \dr{rad}
but other than that not to alter the behaviour
of $\beta$ and $c_0$.  Furthermore in all parts of the plane $c_1
\rightarrow 0$ (see figures \dr{expup} and \dr{expcrash}).
Finally, we observe that when $\beta \rightarrow
\infty$ it grows with time as $\beta \sim t^{1-K}$.  Correspondingly
$c_0$ and $c_1$ tend to
zero exponentially; $c_0, c_1 \sim \exp \left( \frac{-\GGmu
t^{1-K}}{1-K} \right)$.  On the other hand when $\beta \rightarrow 0$
it decreases as $\beta(t)^{1/2} \sim \beta(t_I)^{1/2} - A \ln (t/t_I)$
where $A
= 3\sqrt{\pi} K/8 \sqrt{a}$.

\begin{figure}[ht]
\centerline{\hbox{
\psfig{figure=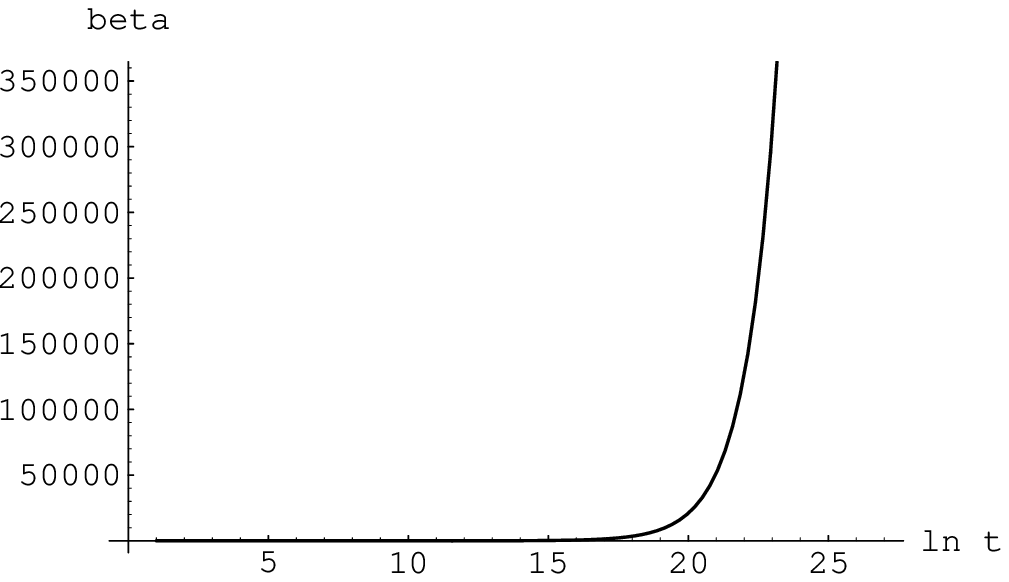,height=5.5cm,width=5.5cm}
\psfig{figure=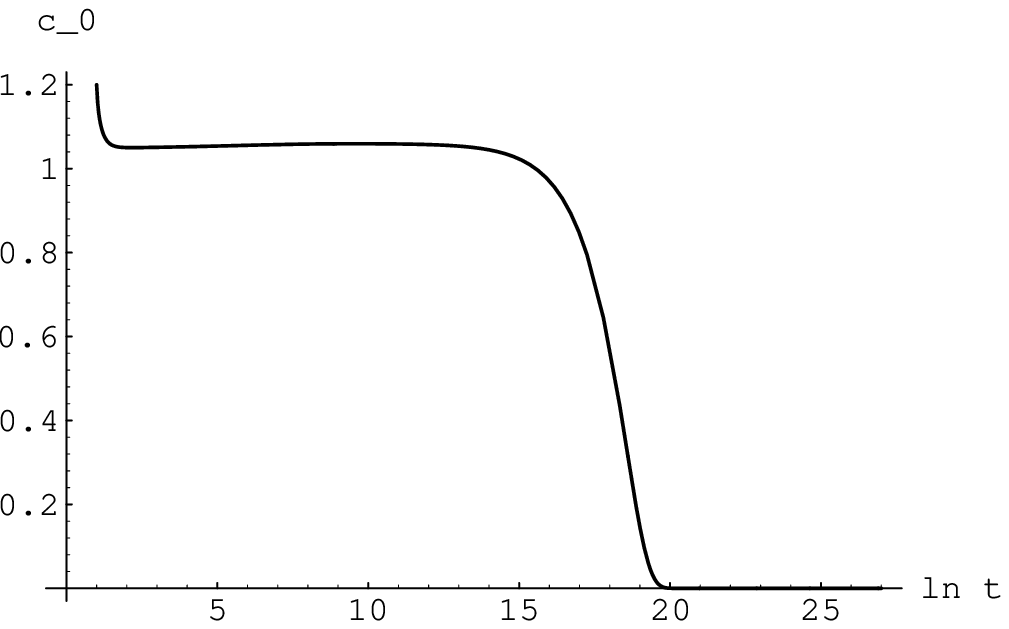,height=5.5cm,width=5.5cm}
\psfig{figure=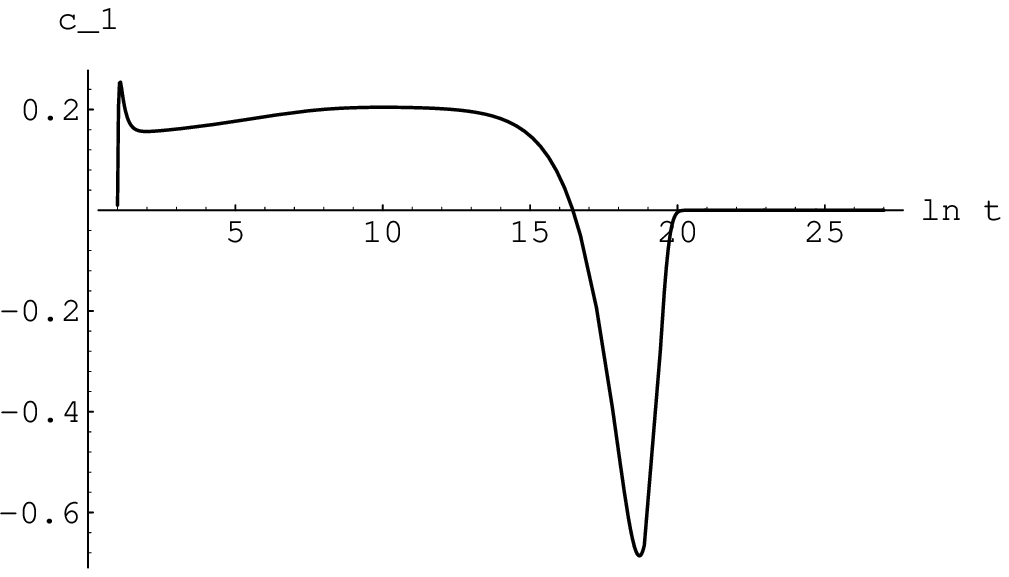,height=5.5cm,width=5.5cm}}}
\caption[Evolution of $\beta$, $c_0$ and $c_1$ in the 
radiation era.]{Evolution of $\beta$, $c_0$ and $c_1$ in the 
radiation era
according to equations (\dr{bcompeqn})-(\dr{c1compeqn}).  Here
$\beta(1) = 0.05$, $c_0(1)=1.2$, $c_1(1)=0.01$, $\GGmu = 10^{-3}$ and
the other parameters
are as in the previous plots. }
\label{expup}
\end{figure}

\begin{figure}[ht]
\centerline{\hbox{
\psfig{figure=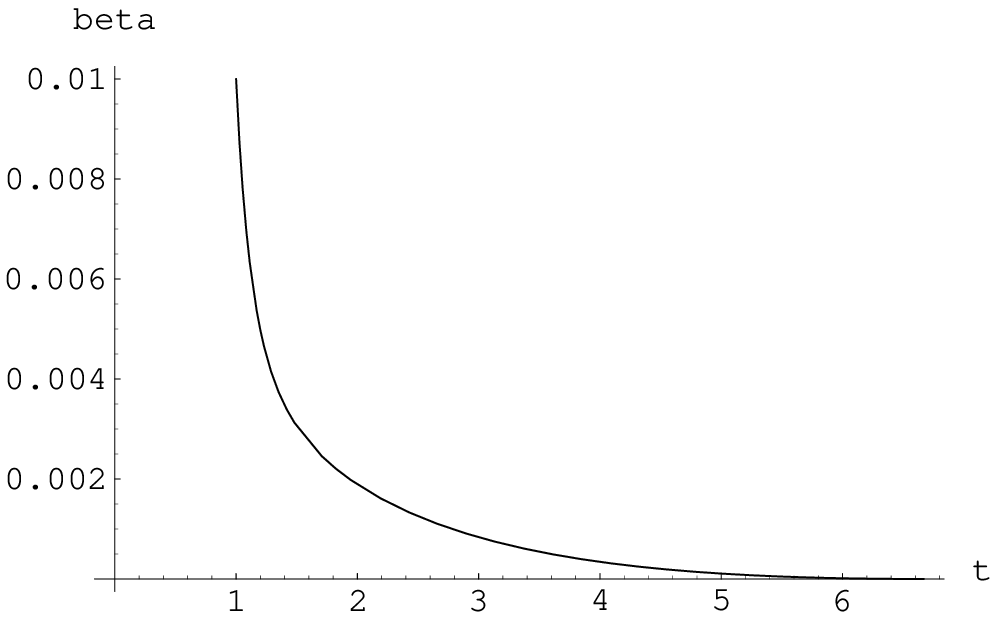,height=5.5cm,width=5.5cm}
\psfig{figure=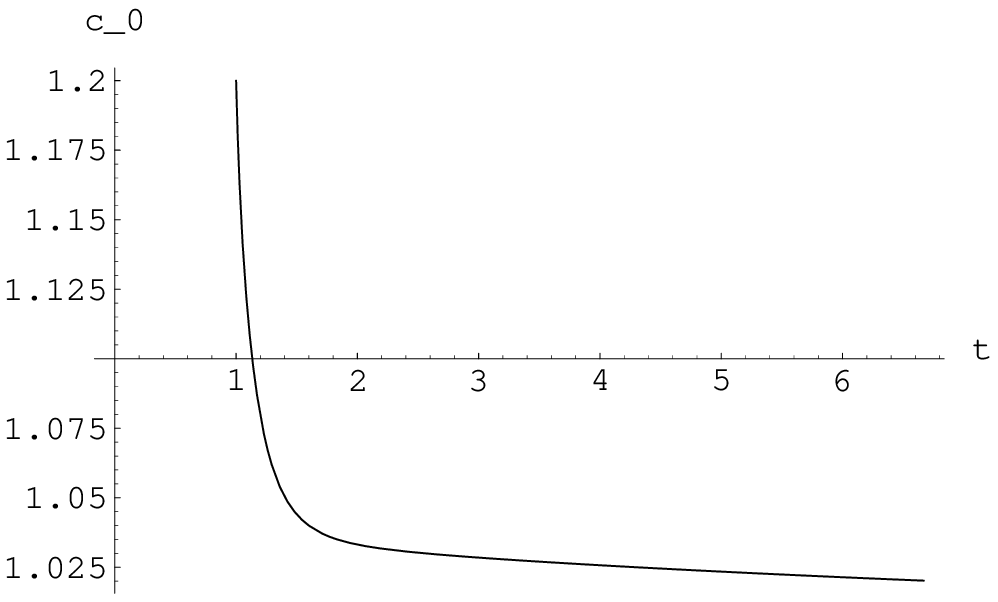,height=5.5cm,width=5.5cm}
\psfig{figure=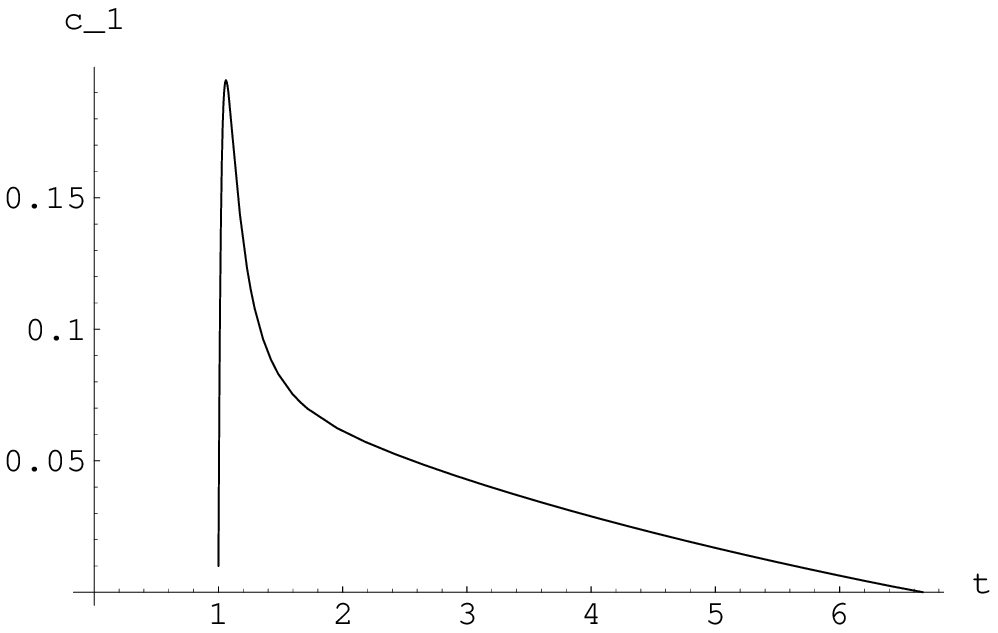,height=5.5cm,width=5.5cm}}}
\caption{Evolution of $\beta$, $c_0$ and $c_1$ in the radiation era with
$\beta(1) = 0.01$, $c_0(1)=1.2$, $c_1(1)=0.01$. The other parameters
are as in the previous plots.}
\label{expcrash}
\end{figure}

\subsubsection{Matter era}

In the matter era, we have argued that there is no scaling solution.
Equation (\dr{cexp}) though can still be written
in the form (\dr{star}), where now
\be
8 \sqrt{a} (\chi {c}_0^{crit} -  \tchi) = - \left( \half -
\frac{3K}{2} \right) < 0
\dle{starstarmat}
\ee
since $K \, \lsim \, 0.1$.  In this case therefore $c_0^{crit}$ may be
positive or negative.  We see that $c_0^{crit} > 0$ if and only if
$\sqrt{a} > \frac{1}{8 \tchi}\left( \frac{1}{2} - \frac{3K}{2}
\right)$.  As in the radiation era, $c_0 \rightarrow {c_0^{crit}}$.
On the other hand, if $c_0^{crit}<0$ then $c_0 \rightarrow 0$.

Let us first consider the case in which $c_0(t_I)={c_0^{crit}} >
0$.  Then since $(\chi {c}_0^{crit}-  \tchi)$ is negative, (\dr{betaexp})
shows that $\dot{\beta}$ is always positive and so $\beta$ grows.  
It increases
until $\beta \simeq 1/\GGmu$ at which stage $c_0 \rightarrow 0$.  We
therefore see that in
contrast to the radiation era, the case for which
$c_0(t_I)={c_0^{crit}}$ always leads to disappearing loops.

However, independently of the value of $c_0^{crit}$, it is possible
to find values of $c_0(t_I)$ for which $\beta
\rightarrow 0$:  from (\dr{betaexp}) this occurs when
$c_0(t_I)$ is very large (see figure \dr{radmat}).  The corresponding 
region of
parameter space though is much smaller than in the radiation era, so
the loops are much more likely to disappear in the matter era.  It
seems that rather extreme initial conditions are required to ensure
$\beta \rightarrow 0$.
The overall situation is summarised in figures
\dr{radmat}--\dr{radmat2}.
In all cases, analysis of equations (\dr{bcompeqn})--(\dr{c1compeqn})
shows that $c_1 \rightarrow 0$.


\begin{figure}[htb]
\setlength{\unitlength}{0.1cm}
\begin{picture}(0,0)
\put(128,-78){\makebox(0,0)[lb]{\large $\beta(t_{I})$}}
\put(32,-16){\makebox(0,0)[lb]{\large $c_0(t_I)$}}
\end{picture}
{\centerline {\epsffile{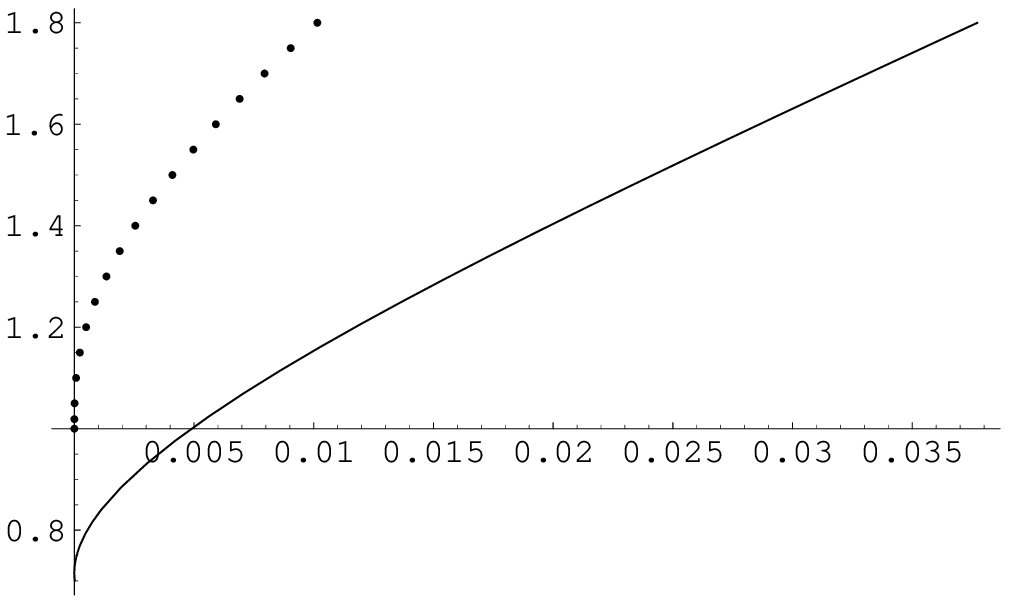}}}
\caption[The $(c_0,\beta)$ plane at the initial time $t=t_{I}$.]{The 
$(c_0,\beta)$ plane at the initial time $t=t_{I}$.  The
solid line is the same curve as in figure \dr{rad} for the radiation
era and $a=1$.  The dotted points are related to the matter era (again
$a=1$).
Here, in contrast to the radiation era, there is no possible scaling
solution for all times:  though $c_0$ tends to a constant as long as
${a^{1/2}} > \frac{1}{8 \tchi}\left( \frac{1}{2} - \frac{3K}{2}
\right)$, there is no solution with $\dot{\beta} = 0$ as $t
\rightarrow \infty$.  Therefore the points on the
dotted line do {\em not} correspond to (unstable) scaling solutions as
they did in the radiation era.
Instead this curve just marks the boundary between the regions for
which $\beta \rightarrow 0$ (on and above the dotted line), and $\beta
\rightarrow \infty$.  In all cases addition of $c_1$ shows that $c_1
\rightarrow 0$ always also in the matter era.}
\label{radmat}
\end{figure}


\begin{figure}[ht]
\centerline{\hbox{
\psfig{figure=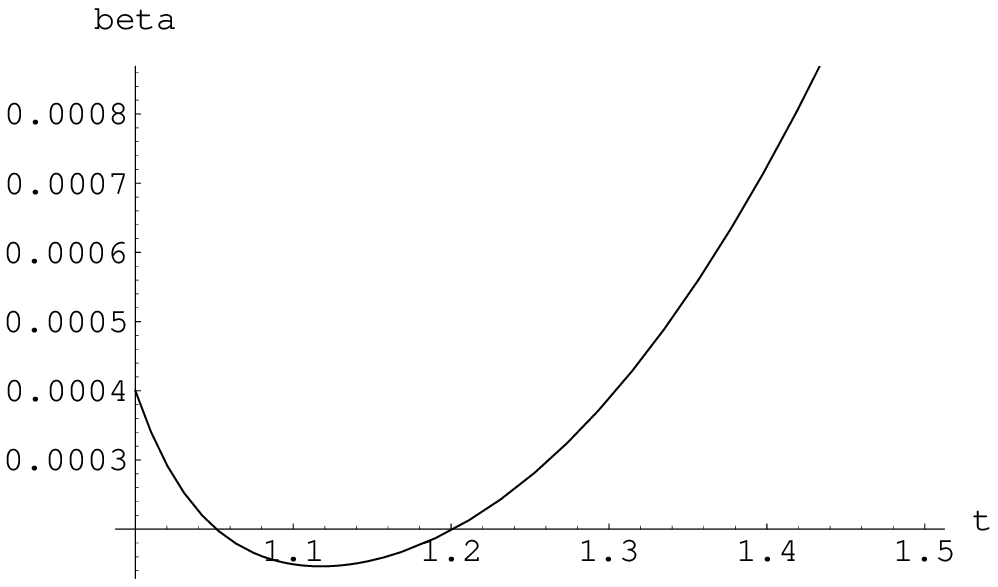,height=5.5cm,width=5.5cm}
\psfig{figure=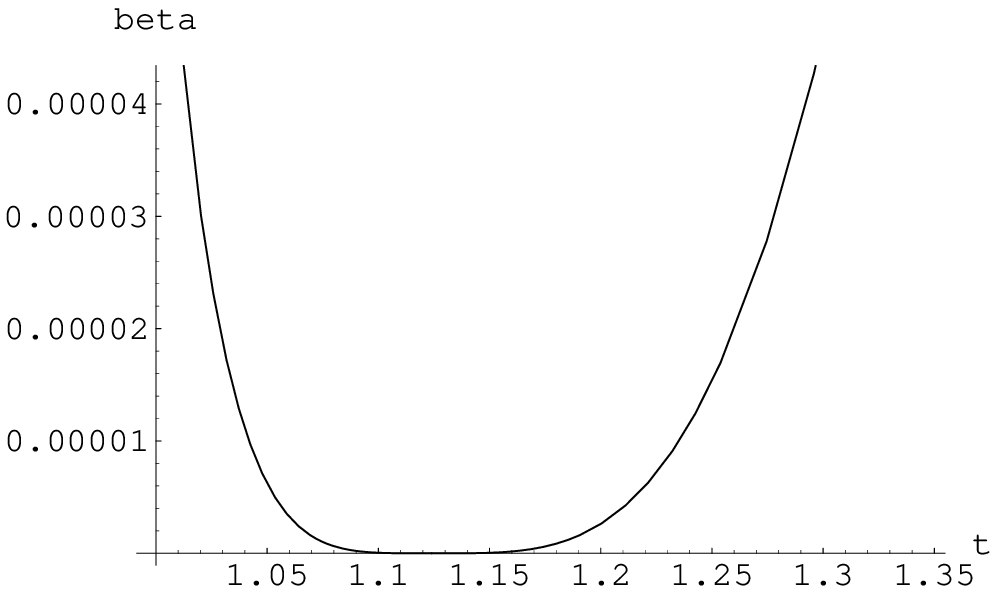,height=5.5cm,width=5.5cm}
\psfig{figure=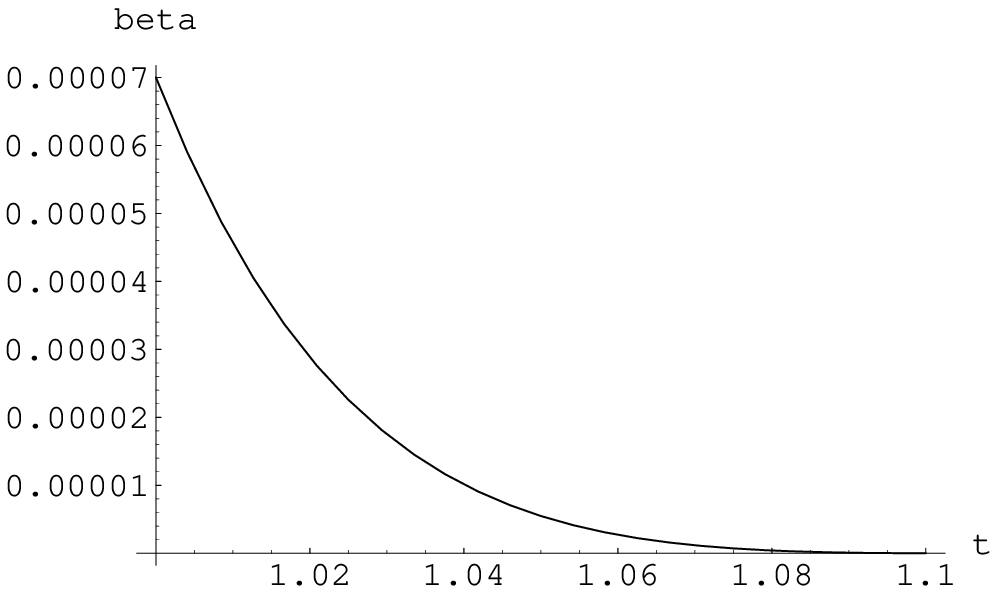,height=5.5cm,width=5.5cm}}}
\caption[Evolution of $\beta$, $c_0$ and $c_1$ in the matter 
era.]{Evolution of $\beta$, $c_0$ and $c_1$ in the matter era
according to equations (\dr{bcompeqn})-(\dr{c1compeqn}).
In all of these figures $c_0(t_I)=1.1$ and $a=1$.  From
figure \dr{radmat}, the value of
$\beta(t_I)$ which cuts off between $\beta \rightarrow 0$ and $\beta
\rightarrow \infty$ is $\beta^{crit}
= 0.0000725$.
In the first figure $\beta(t_I) = 0.0004 > \beta^{crit}(t_I)$ and
$\beta \rightarrow \infty$.  In the
second $\beta(t_I) = \beta^{crit}(t_I)$, whilst in the third
$\beta(t_I) = 0.00007 < \beta^{crit}(t_I)$.}
\label{nonup2}
\end{figure}
\begin{figure}[ht]
\centerline{\hbox{
\psfig{figure=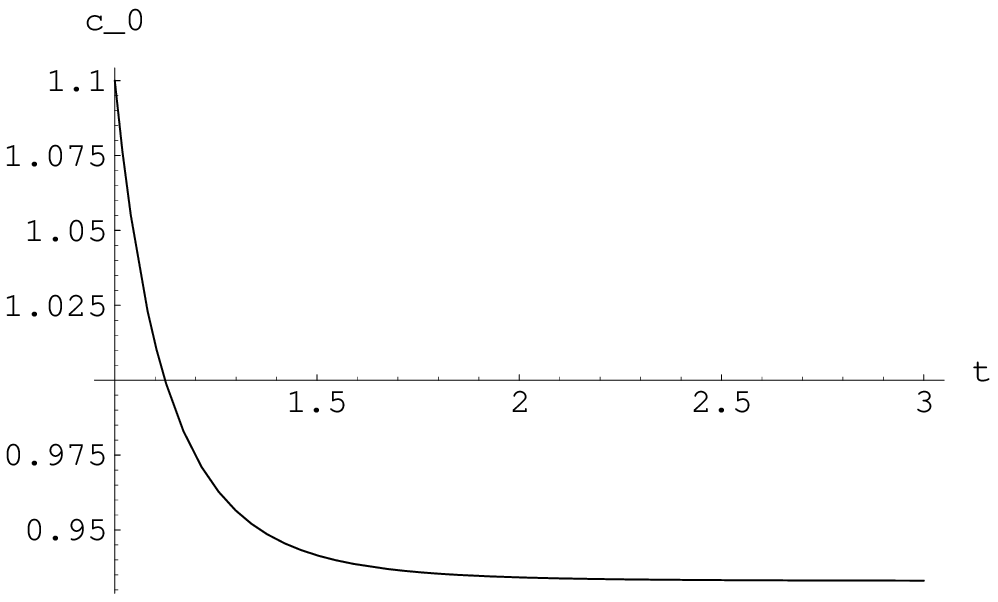,height=5.5cm,width=5.5cm}
\psfig{figure=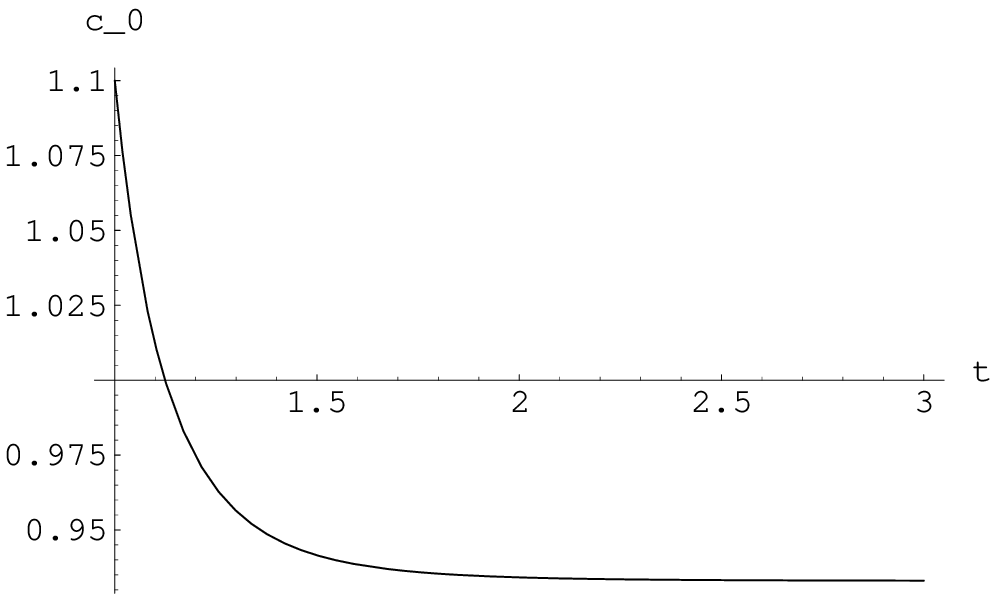,height=5.5cm,width=5.5cm}
\psfig{figure=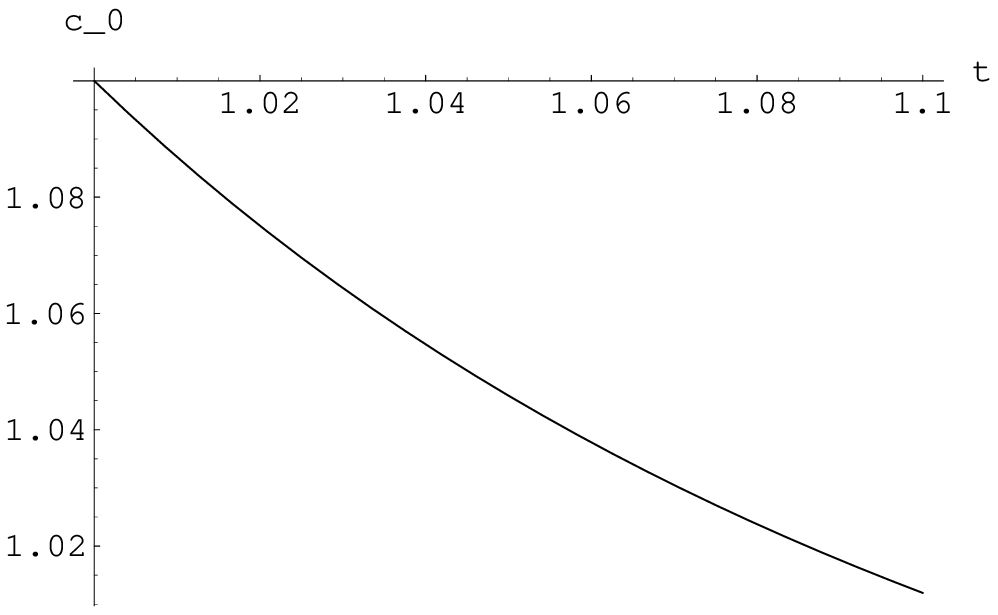,height=5.5cm,width=5.5cm}}}
\caption{The evolution of $c_0$ corresponding to the previous figure.
Thus $c_0(t_I)=1.1$, and we have the different values of $\beta(t_I) $
as given above.  In the first two figures $c_0$ will eventually tend to
zero when $\beta$ gets very large.   }
\label{radmat2}
\end{figure}

\section{Search for power-law solutions}\dle{inf}

We have seen that both in an expanding as well as in a non-expanding
universe, if the initial value
of $c_0$ is large enough the solution will evolve towards $\beta=0$.
What happens when this point is reached?

In section \dr{full}, we performed a Taylor series expansion in
inverse powers of $a/z$, looking for a solution of the form
(\dr{hope4}).  However, this expansion is invalid when
$\beta=0$.  There the integral (\dr{term2newB}) representing Term
2 is no longer an analytic function in the neighbourhood of
$z=\infty$.  It can be expressed in terms of hypergeometric
functions, and expanded in powers of $(a/z)$ to give
 \be
{\rm Term2} = {\tilde\chi\over\tilde\xi_c^{3/2}}
\sum_{n=0}^\infty \frac{c_n}{z^2}\left[\bullet+\bullet{1\over z^{1/2}}
+\bullet{1\over z}+\cdots\right]
\qquad(z\gg a).
\dle{expnonanaly}
 \ee
The appearance of half-integral powers implies that the
ansatz (\dr{hope4}) fails at this point.

Physically, this failure is unsurprising.  We know that in
equilibrium in a non-expanding universe, there is a phase
transition, the Hagedorn transition, at the point $\beta=0$
\dc{Edandco,MT,Ant2}.  There is a critical value of the energy
density, $E_{crit}$.  If we start with a low density,
$E<E_{crit}$, the system will tend towards an equilibrium
state described by fixed values of $\beta$ and $c_0$, where
the curves cross in figure \dr{phasenonexp}.  For large density,
however, the equilibrium state contains not only a loop
distribution with $g(z)\propto z^{-5/2}$ (i.e., $\beta=0$) but
also a single infinite string.  The normalisation of the loop
distribution (i.e.\
$c_0$) is independent of the density; as the density increases more and
more of the length goes into infinite string.

Our equations as they stand do not explicitly allow for
infinite strings.  The transition is signalled by
the fact that the equations break down at that point.  If we
start in a non-expanding universe, with $E>E_{crit}$, the
solution will evolve toward $\beta=0$.  It will reach that
point in a finite time.  However the resulting value of $c_0$
exceeds the maximum value for which a stable solution is
possible, $c_0^{crit}$.  What we believe happens therefore is
that the excess energy is transferred into the form of long
strings, leaving a scale invariant loop distribution characterised by
$c_0=c_0^{crit}$.

At first sight it may perhaps seem surprising that a pure
loop distribution can evolve to create infinite strings.
However, it is not difficult to see that this can indeed
happen.  Creating an infinite string obviously requires an
infinite number of intercommuting events, but the number per
unit length or unit volume is finite.  If we start with two
infinite sinusoidal strings approaching one another
transversely, they can break up into an infinite set of loops
(see figure \dr{join}).  Conversely, an infinite set of loops can
collide to form a pair of infinite strings.  Indeed if we
start with a sufficiently large density of loops, we must
expect that mutual collisions will lead to a percolating
system, i.e., to infinite strings.
\begin{figure}[htb]
{\centerline {\epsfxsize=6in \epsfysize=5in \epsffile{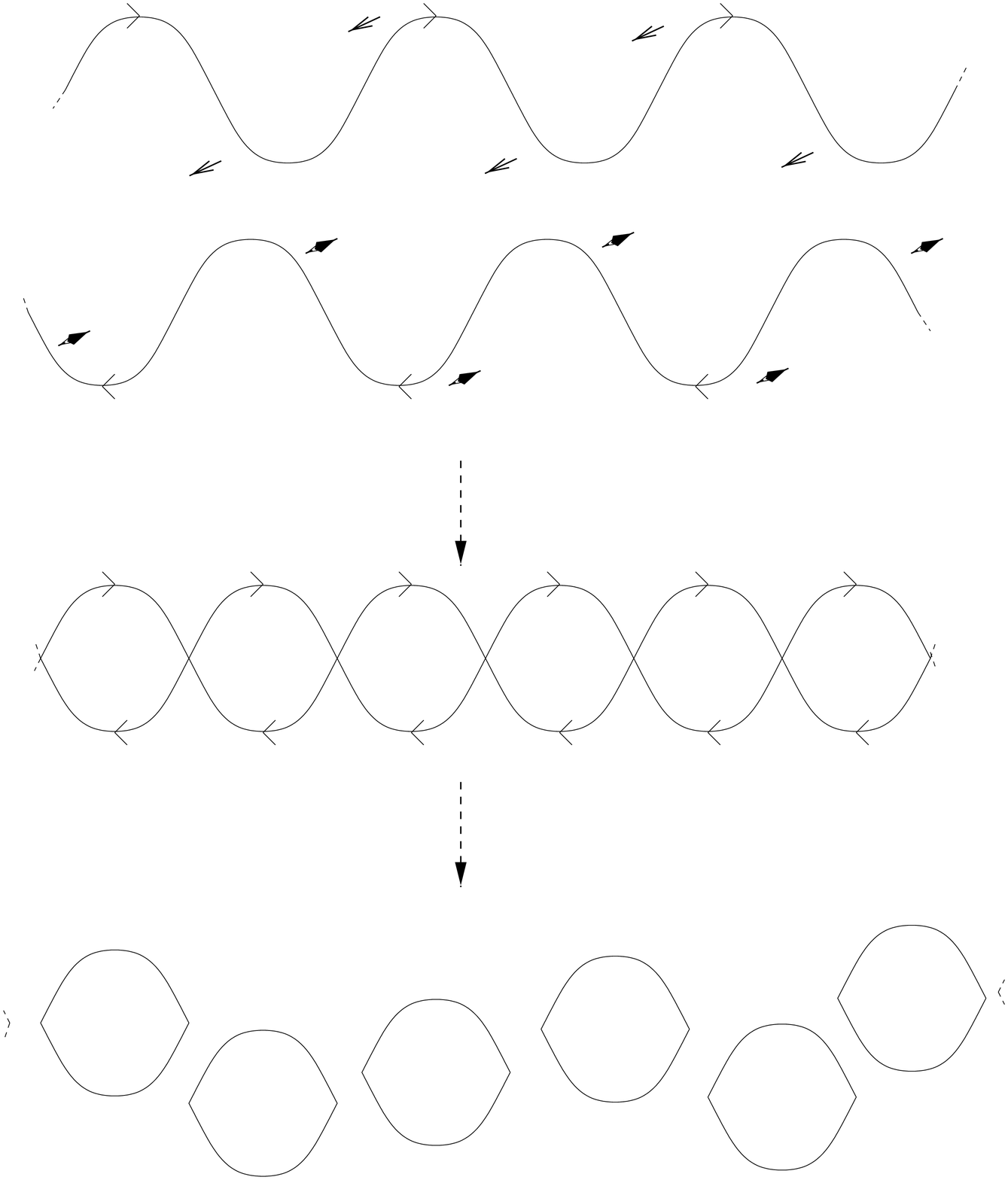}}}
\caption{Two sinusoidal infinite strings travel towards each other
generating an infinite set of loops.}
\label{join}
\end{figure}

It should be possible to augment our equations by allowing
also for a certain density of infinite string.  We have not
attempted to do this in detail (because of certain technical
difficulties).  However, qualitative examination of the
problem suggests that infinite string could not be formed or
survive if the loop distribution corresponded to $\beta>0$;
in such a case the probability of creating loops from
infinite strings would be infinitely greater than that of
creating infinite strings from loops.  For $\beta=0$ the
equations would appear to be consistent only when
$c_0=c_0^{crit}$.  Thus infinite strings must always be
accompanied by the invariant loop distribution known to be
present in equilibrium.

We believe essentially the same results would apply in a
radiation or matter dominated universe.  If the initial string
distribution consists only of loops, but with a density above
some critical value, they will evolve towards a distribution
in which infinite strings are present ($\beta = 0$).  However, such an
initial distribution is perhaps rather unlikely: if the
density is large enough, then in all probability the initial
distribution itself would already contain infinite strings.
Physically, the more interesting situation is the one in
which the initial density is below the critical value, so
that the distribution contains only loops for all times.  It will then
evolve as described in the previous sections and the loops will
disappear in the matter era.

\section{Conclusions}\dle{conc}

We have set up a model for the non-equilibrium evolution of a network
of relativistic cosmic string containing {\em only} loops,
taking into account the possibility
that loops can join to form longer loops, and that a loop can 
self-intersect.  The effects of gravitational radiation, redshift and
expansion on the length of a loop were also considered.

In a non-expanding universe we showed that the model conserves the
total energy density in loops, $E \propto \int_0^{\infty} \nlt \k
\dk$, as required for consistency.  Furthermore, under the assumption
that the two length scales $\bxi$ and $\zeta$ characterising the loops
are constant in time, we were able to show that if $E$ is less than
some critical energy density $E_{crit}$, the loops reach a {\em stable}
equilibrium distribution of the form $n(\k) = c_0^{crit} \k ^{-5/2} e^{-\beta
 \k}$.  The value of $E_{crit}$ is determined by $\beta = 0$ and $c_0^{crit} =
\tchi/\chi$; that is by the ratio of two dynamical parameters in the
model.  For $E>E_{crit}$ we argued that $\beta \rightarrow 0$ and
hence that a scale invariant distribution of loops is formed.
Furthermore, by combining the fact that energy is conserved with an
extension of the model to include infinite strings, we argued
that when the point $\beta = 0$ is reached a distribution of loops of
the form $\nlt = c_0^{crit} \k^{-5/2}$ is formed with the rest of the
initial energy density going into infinite strings.  These results are
consistent with those of statistical mechanics.

In an expanding universe we showed that the situation for the evolving
loop network is very different.  In a
radiation dominated universe a scaling solution was observed to exist, but
it was seen to be an {\em unstable} one.  Also, depending on the
initial energy density in the loop network (figure \dr{rad}), we saw
that either $\beta \rightarrow \infty$ so that the loops disappear, or
$\beta \rightarrow 0$.  In a matter dominated universe, there is no
possible scaling solution.  Given the analysis of section \dr{sscalvar} we
conclude that in an expanding universe, the network may evolve in the
following way --- depending on the initial distribution of loops formed
at the phase transition:
\begin{itemize}
\item  If the energy density in loops is small, the loops
disappear both in the radiation and the matter eras.
\item  If the energy density in loops is greater than a critical
value then $\beta \rightarrow 0$ in the radiation era, and we may
expect the following possibilities:
\begin{enumerate}
\item The point $\beta = 0$ is reached still in the radiation era or
conceivably in the matter era, in
which case infinite strings are formed with a scale invariant
distribution of loops.  The rate equation then breaks down and the
effect of infinite strings need to be included (see for example
\dc{ACK}).
\item If $\beta$ does not reach zero we expect the loops to disappear
in the matter era.
\end{enumerate}
\end{itemize}
We thus believe that the only way in which the loop network can evolve
so as to stay as a loop network (that is one not containing infinite
strings) is for the loops to disappear in the matter era.  This
scenario may be interesting from the view point of structure formation
from cosmic strings \dc{James}.

Throughout this work we have assumed that in an expanding universe
$\bxi$ and $\zeta$ scale.  A full analysis would not make this
assumption but itself derive equations for $\bxi(t)$ and $\zeta(t)$,
though this would be very difficult ({\em c.f.}\ \dc{ACK}).  Also, we
have analysed the evolution of specific initial distributions
$\gzt
\propto z^{\alpha} e^{-\beta z}$ and not arbitrary ones which are
outside the scope of our analytical work.  However, given that results
from string
statistical mechanics and simulations of cosmic
string formation predict initial loop distributions of the type we
have analysed, we believe we have studied the most relevant
case.  The study of arbitrary distributions requires a
numerical approach \dc{Joao}.

There are a number of directions in which we would like to extend this
work.  Firstly we hope to study carefully the effect of the transition
from radiation to matter domination, and hence to clarify which of the
above outcomes is the most likely.  We also hope to explore the
cosmological consequences of the
evolution of such a network and calculate for example the
gravitational radiation emitted.  If infinite strings are proved to
exist then we hope to add their effect to this model, for
example by combining it with the 3-scale model \dc{ACK}.

\section*{Acknowledgements}

D.A.S.\ thanks Tim Evans for numerous useful discussions, and also
Jo\~{a}o Magueijo and Ray Rivers.
E.J.C.\ and D.A.S.\ are supported by P.P.A.R.C.\, and D.A.S.\
is a member of Girton College, Cambridge.
This work was supported in part
by the European Commission under the Human Capital and Mobility
programme, contract number CHRX-CT94-0423 and is also the
result of a Scientific Network supported by the European Science
Foundation.

\typeout{--- No new page for bibliography ---}


\begin{thebibliography}{99}


\bibitem{James} A.Albrecht, R.A.Battye and J.Robinson, Phys.\ Rev.\
Lett.\ {\bf 79} (1997) 4736.

\bibitem{Tom} T.W.B.Kibble, J.\ Phys.\ A: Math.\ Gen.\ {\bf
9} (1976) 1387.

\bibitem{HiKi} M.Hindmarsh and T.W.B.Kibble, Rep.\ Prog.\  Phys.\ {\bf 58}
(1995) 577.

\bibitem{ViSh}  A.Vilenkin and P.Shellard,
Cosmic Strings and Other Topological Defects (Cambridge University
Press, Cambridge, 1994).

\bibitem{VV} T.Vachaspati and A.Vilenkin, Phys.\ Rev.\ {\bf D30}
(1984) 2036.


\bibitem{AT} A.Albrecht and N.Turok, Phys.\ Rev.\ Lett.\ {\bf
54} (1985) 1868.


\bibitem{BB} D.P.Bennett and F.R.Bouchet, Phys.\ Rev.\ Lett.\ {\bf
63} (1990) 2776.

\bibitem{AS} B.Allen and E.P.S.Shellard, Phys.\ Rev.\ Lett.\
{\bf 64} (1990) 119.


\bibitem{ACK} Daren Austin, E.J. Copeland and T.W.B. Kibble,
Phys.\ Rev.\ {\bf D51} (1995) 2499.


\bibitem{CJAP}  C.J.A.P.Martins and E.P.S.Shellard, Phys.\ Rev.\
{\bf D54} (1996) 2535-2556.




\bibitem{James2} A.Albrecht, R.A.Battye and J.Robinson, Structure
formation by cosmic strings with a cosmological constant, {\tt
astro-ph/9711336} (1997).


\bibitem{Avelino} P.P.Avelino, E.P.S.Shellard, J.H.P.Wu, B.Allen,
Cosmic string-seeded structure formation, {\tt astro-ph/9712008} (1997).


\bibitem{Kaj} K.Kajantie, M.Laine, K.Rummukainen and
M.Shaposhnikov, Nucl.\ Phys.\ {\bf B493} (1997) 413.


\bibitem{RayGlk} G.Karra and R.J.Rivers, Phys.\ Lett.\ {\bf B414}
(1997) 28-33.



\bibitem{KarlHi} M.Hindmarsh and K.Strobl, Nucl.\
Phys.\ {\bf B437} (1995) 471.


\bibitem{Julian} J.Borrill, Phys.\ Rev.\
Lett.\  {\bf 76} (1996) 3255.

\bibitem{ScVi} R.J.Scherrer and A.Vilenkin, Phys.\ Rev.\ {\bf D56}
(1997) 647.

\bibitem{AndyJames} J.Robinson and A.Yates, Phys.\ Rev.\ {\bf D54}
(1996) 5211.



\bibitem{AndyTom} A.Yates and T.W.B.Kibble, Phys.\ Lett.\ {\bf B364}
(1995) 149.

\bibitem{MarkGray} G.Vincent, N.D.Antunes and M.Hindmarsh,  Phys.\
Rev.\ Lett.\ {\bf 80} (1998) 2277.



\bibitem{Vach} A.A.\ de Laix and T.Vachaspati, On Random Bubble
Lattices, {\tt hep-ph/9802423}, (1998).

\bibitem{Ant2} N.D.Antunes, L.M.A.Bettencourt and M.Hindmarsh, Phys.\
Rev.\ Lett.\ {\bf 80} (1998) 908.



\bibitem{Pedro} P.Ferreira, Ph.D. thesis, University of London,
(1995).





\bibitem{LL} L.D.Landau and E.M.Lifshitz, Physical Kinematics
(Pergamon Press, Oxford, 1981).

\bibitem{Bern} J.Bernstein, Kinetic Theory in the Expanding
Universe (Cambridge University Press, 1988).

\bibitem{KT} E.W.Kolb and M.S.Turner, The Early Universe
(Addison-Wesley, 1994).




\bibitem{Thesis} D.A.Steer, Ph.D. thesis, University of London,
(1997).


\bibitem{MGM} M.Hindmarsh, M.Sakellariadou and G.Vincent, Phys.\
Rev.\ {\bf D56} (1997) 637.

\bibitem{HGr} M.Hindmarsh, Phys.\ Lett.\ {\bf B251} (1990) 28.


\bibitem{Joao} J.Magueijo and D.A.Steer,
in progress.






\bibitem{MT} D.Mitchell and N.Turok, Phys.\ Rev.\ Lett.\ {\bf 58}
(1987) 1577.

\bibitem{Edandco} E.J.Copeland, D.Haws, S.Holbraad and R.Rivers,
in:  The Formation and Evolution of Cosmic Strings, ed. G.Gibbons,
S.Hawking, T.Vachaspati (Cambridge University Press, 1990) p.35.

\bibitem{SakVil} M.Sakellariadou and A.Vilenkin, Phys.\ Rev.\ {\bf
D37} (1988) 885.






\end{thebibliography}
\end{document}